\documentclass[aps,prd,showpacs,preprintnumbers,superscriptaddress,nofootinbib]{revtex4}
\usepackage[dvipdfmx]{graphicx}
\usepackage{bm,latexsym,amsmath,amssymb,amsfonts}

\begin{document}
\title{Informational Theory of Relativity}
\author{Akio~Hosoya}
\email[Email: ]{ahosoya@phys.titech.ac.jp, ahosoya.bongo@gmail.com}
\affiliation{Department of Physics, Tokyo Institute of Technology, Tokyo 152-8551, Japan}

\author{Shunsuke~Fujii}
\email[Email: ]{fujii@tokyo-ct.ac.jp}
\affiliation{Department of General Education, National Institute of Technology, Tokyo College, Tokyo 193-0097, Japan}

\today

\begin{abstract}
 Assuming the minimal time to send a bit of information in the
Einstein clock synchronization of the two clocks located at different positions, we introduce the extended metric to the information space. This modification of relativity changes the red shift formula
keeping the geodesic equation intact. Extending the gauge symmetry hidden in the metric to the 5-dimensional general invariance, we start with the Einstein-Hilbert action in the 5-dimensional space-time. After the  4+1 decomposition, we obtain the effective action which includes the Einstein-Hilbert action for gravity, the Maxwell-like action for the velocity field and the Lagrange multiplier term which ensures the normalization of the time-like velocity field. As an application, we investigate a solution of the
field equations in the case that a 4-dimensional part of the extended metric is spherically symmetric, which exhibits Schwarzschild-like space-time but with the minimal radius. As a discussion we present a possible informational model of synchronization process which is inherently stochastic. The model enables us to interpret the information quantity as a new spatial coordinate. 

\end{abstract}

\newcommand{\eq}[1]{Eq.~{(\ref{#1})}}
\newcommand{\fig}[1]{FIG.~{\ref{#1}}}
\newcommand{\bea}{\begin{eqnarray}}
\newcommand{\eea}{\end{eqnarray}}
\newcommand{\bc}{\left ( \begin{array}{c}}
\newcommand{\ec}{\end{array} \right )}

\newcommand{\fr}[2]{\frac{#1}{#2}}
\newcommand{\pa}{\partial}

\newcommand{\uf}{{}^{(5)}}


\maketitle

\section{Introduction}

\begin{figure}[h]
  \begin{center}
    \includegraphics[width=0.3\columnwidth,clip]{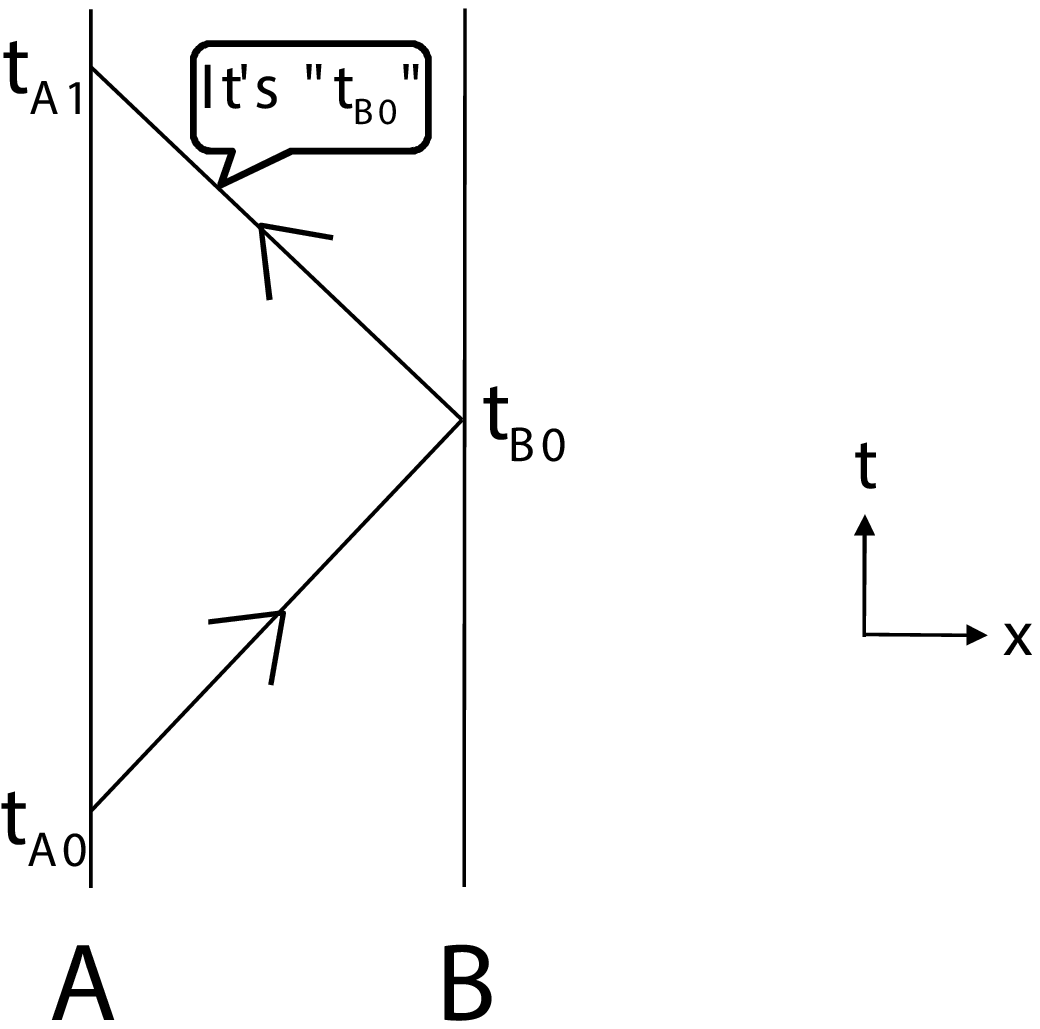}
  \end{center}
  \caption[The Einstein synchronization]{The Einstein synchronization \newline An observer {\it A} (Alice) needs to be told the arriving time $t_{B0}$ measured by an observer {\it B} (Bob) to confirm that the spatially separated clocks {\bf a} and {\bf b} are synchronized.}
  \label{fig:STwithInfo}
\end{figure}

 In the seminal  paper on the special relativity published in 1905~\cite{Einstein1905}, Einstein considered a thought experiment for two spatially separated clocks to synchronize by exchanging light signals back and forth between them.
An \emph{observer A} (Alice) with a clock {\bf a} can confirm that the clocks {\bf a} and {\bf b} are synchronized if she finds that
\bea
	\fr{t_{A0}+t_{A1}}{2}=t_{B0}  \label{clkAB}
\eea   
holds. The first light signal starts from Alice at the time $t_{A0}$ as the clock {\bf a} indicates, reaches the other \emph{observer B} (Bob) at the time $t_{B0}$ measured by the clock {\bf b} and then the information of the numerical value $t_{B0}$ is sent back to Alice by the light signal, who receives the information of $t_{B0}$ at the time $t_{A1}$. By this communication Alice obtains all of the necessary data $t_{A0},t_{B0}$ and $t_{A1}$ to check the synchronization criterion (\ref{clkAB}). (See \fig{fig:STwithInfo})
 \footnote{For Bob also to convince himself the synchronization, Alice has to send the information $``t_{A1}"$ to Bob, which is received at $``t_{B1}"$. If the relation $\fr{t_{B0}+t_{B1}}{2}=t_{A1}$ holds, Bob confirms
the synchronization.}
Einstein then applied the principle of special relativity to the synchronization\footnote{Einstein's synchronization in the rest frame can be extended to all the clocks, which defines the constant $t$ surface in the space-time, while the synchronization in the moving frame defines the another constant time surface. Although in the both frames the synchronization can be carried out, the resultant constant time surfaces are different which is known as the relativity of simultaneity. One should not confuse the two concepts the synchronization and the simultaneity. } of the two separated clocks in each inertial frame and arrived at the Lorentz transformation between the two inertial frames as reproduced in Appendix {\ref{app:A}}.

Note that the argument made by Einstein here is {\it operational}, i.e., the each statement in the argument can be verified by experiments in principle. In general one of the merits of 
operational argument is that the  assumption we made is physically unambiguous. Then we can clearly see that Einstein tacitly assumed that the time duration for sending information of $t_{B0}$ is zero as an idealization.

Let us instead  think about the more realistic possibility by taking into account the time duration assuming that we need the minimum time duration $\tau$ to send a unit of information. This operationally introduced assumption makes special relativity accommodate with the information theory~\cite{Shannon}, and enables us to extend the ordinary space-time in the direction of information space, the exact meaning of which will become clearer in especially Sec.\ref{sec:orgin_sigma}. With this set up, both of the information theory and the special relativity can be further extended into the informational theory of general relativity.

In the standard relativity, an event is specified by the time $t$ and the position $x$ at which the event occurred. In our present work, we demand that the communication time duration is minimal in the spirit of Fermat's principle. For this purpose we maximally shorten the exchanged data on the basis of Shannon's theorem of optimal compression. 
That is, the original $n$ bits of information is compressed to the the $\sigma n$
 bits where $\sigma $ is the maximal compression rate.
As is well known, the data compression is the central concept of Shannon's information theory\cite{Shannon}. Note that the Shannon information entropy is the average of the compression rate $\sigma$. We promote the compression rate $\sigma$ as an extra coordinate in addition to the ordinary space-time coordinate $(t,x)$. An event is therefore specified by space-time information coordinates $(t,x,\sigma)$.
This seems natural, because we always ask when, where and what about the event.
Let the two events be $(t_1,x_1,\sigma_1)$ and  $(t_2,x_2,\sigma_2)$ and let us consider the communication of signals between observers at two different space-time points. If these events correspond to the sending and the receiving of the signals, it is necessary to satisfy $\sigma_1 n=\sigma_2 n$, because the information of the sender is copied by the receiver after the signal processing and is shared between them. The Einstein synchronization is a special case that the processing is instantaneous. ($\tau =0$)

Now let us shift the view point from the operational description of informational space-time to the usual geometrical point of view. We can generalize ordinal space-time into the informational space-time by thinking of the distance between the two events as the sum of the ordinary pseudo-Riemannian space-time metric and the informational distance $(\sigma_1-\sigma_2)^2$ with some weight. The latter corresponds to the Fisher metric in the information theory. For further discussion on the meaning the variable $\sigma$, we present a stochastic model of the synchronization of all the clocks in a bounded domain of the universe in Sec\ref{sec:orgin_sigma}.

At this stage we emphasize that 
the time and the space as the argument of the metric tensor field are the readings of the clock and the measure nearby the observer so that the time delay of the informational signal should be taken into account to obtain the real time ($\tau=0$\;;the  original time introduced in special relativity) which can be expressed  as the observed arrival time of the information subtracted by the time delay $\tau \sigma$. 
On the other hand the causality is defined in terms of the true coordinates of space-time. The difference of the observed and true coordinates is proportional to the informational coordinate $\sigma$.


 The organization of the present work is the followings.
 In section \ref{sec:intro} we operationally develop the basic idea of the informational theory of relativity (ITR) starting with the Einstein synchronization of the two spatially separated clocks and arrive at the metric with the velocity field as a new ingredient.  On the basis of the metric given in section \ref{sec:intro}, we check the classic tests of general relativity
in section \ref{sec:classic} to find that the only deviation from the conventional general relativity is the red-shift formula in subsection \ref{sec:reds}, since the geodesic equation remains intact as shown in subsection \ref{sec:geo}.  The section \ref{sec:symmAction} is divided into the three subsections. In subsection \ref{sec:symm}, we point out a gauge invariance of the
metric, which is a part of the 5-dimensional coordinate transformation. Assuming the 5-dimensional coordinate transformation invariance as the basic symmetry in subsection \ref{sec:action}, we set up the 5-dimensional Einstein-Hilbert action and then decompose it in (4+1) dimensions to obtain the 4-dimensional action as a sum of the modified Einstein-Hilbert action and the Maxwell-like action for the velocity field as well as the constraint term for it while subsection \ref{sec:ActMat} is for other matter fields. We derive the Euler-Lagrange equations for the metric and the velocity fields in subsection \ref{sec:fieldeq}. In subsection \ref{sec:spherical}, we show a spherically
symmetric solution as an example. Sections \ref{sec:summary} and \ref{sec:discussion} are devoted to summary and discussions.
Especially in the discussion \ref{sec:orgin_sigma}, we give a simple stochastic model for the data compression in our synchronization process.
 We present detailed calculations in Appendix A for Einstein's derivation of the Lorentz transformation, formal tensor calculi in Appendices B through D and Appendix E is specifically for the spherically symmetric solution.

 We follow the notations of the book by Misner, Thorne and Wheeler~\cite{MTW} with the metric signature $(-++++)$.
\section{Informational Relativity}\label{sec:intro}

The position and the time are locally and operationally defined by reading a measure and a clock there.

Let us denote $\tau_0$ as the minimum time required to send one bit of information in the rest frame.
The total time duration $T$ to send the message amounts to $T=\tau _0n\sigma$ with $n\sigma$ being the number of the maximally compressed bits to express the information of time such as ``$t_{\rm B0}$''. We fix the number of bits $n$ for the raw data through out the present paper and write $\tau :=\tau _0n$ so that the total time duaration becomes $T=n\tau_0\sigma=\tau\sigma$. 

\begin{figure}[h]
  \begin{center}
    \includegraphics[width=0.2\columnwidth,clip]{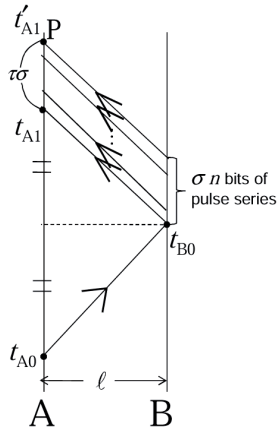}
  \end{center}
  \caption[The time duration of sending $n \sigma$ bits of data]{The time duration of sending $\sigma$ bits of data \newline To send back the light signal of the $\sigma$ bits of information it takes the time duration $\tau \sigma$.}
  \label{fig:duration}
\end{figure}

The received time $t'_{A1}$ includes the time duration
$\tau \sigma $ so that we have to offset  the time duration $\tau \sigma$ from $t'_{A1}$ to infer the true received time $t_{A1}$
\bea
	t_{A1} \longrightarrow t_{A1}:=t'_{A1}-\tau \sigma\;. \label{clkOffset}
\eea 
 Therefore, the two clocks {\bf a} and {\bf b} are confirmed to be synchronized if
\bea
	\fr{t_{A1}+t_{A0}}{2}=\fr{t'_{A1}-\tau \sigma +t_{A0}}{2}=t_{B0} \label{clkAdj}
\eea 
holds in terms of the recorded times $t_{A0}$  and $t'_{A1}$ by Alice and $t_{B0}$ informed by Bob (\fig{fig:duration}).

With the clocks {\bf a} and {\bf b} synchronized by the light signal, the observables  $t'_{A1}, t_{B0}$ and the spatial distance $x$ between A and B are related by 
\bea
	x =c(t'_{A1}-\tau \sigma -t_{B0})\;. \label{clkL}
\eea 
 In order to verify this relation, Alice needs to know the time $t'_{A1}$ when the light signal from the clock {\bf b} arrived, read the information $t_{B0}$ from Bob, and remember the given distance $x$ between the clocks {\bf a} and {\bf b}. After obtaining these three quantities at the same time $t'_{A1}$, the verification of \eq{clkL} is done locally at P in \fig{fig:duration}.\\
 We emphasize that the time and space coordinates $t$ and $x$ are treated as the informational quantities\footnote{Note that by the requirement of the minimum time $\tau$ for sending one bit of data, the space-time coordinates can be rescaled with a unit of $\tau$ or $c\tau$ and then become dimesionless numerical values on the same footing of the information $\sigma$.} on the same footing of the compression rate $\sigma$ obtainable by
 a local observer.  

It is known that this synchronization works provided that the gravitational red shift between the two positions is negligible in the sense that the round trip synchronization is consistent~\cite{Macdonald}. Later we shall come back to this round trip problem from our point of view.
Therefore, at the moment it is safe to apply the Einstein synchronization to two neighboring positions. Assuming $\tau$ is constant, we see that the differential expression of \eq{clkL} is
\bea
	dx=c(dt -\tau d\sigma )\label{clkdx}
\eea
with the coordinates $(x,t,\sigma )$. 


{
So far we have considered the special case that the two observers and clocks are placed at rest in the coordinate system
$(t,x,y,z)$. Let us consider more general case that the two clocks 
are sitting in the frame $(\eta,X,Y,Z)$ moving in the $x$-direction at the velocity $v$ relative to the rest frame with the coordinates
$(t,x,y,z)$. Note that Einstein's synchronization works as far as the observers $A$ and $B$ comove with the clocks ${\bf a}$ and ${\bf b}$ in any inertial frame. To avoid possible confusion we eliminate the observers $A$ and $B$ and adopt an automatic synchronization processing of the clocks $a$ and $b$ by exchanging light signals.

For the two clocks fixed in the frame of the coordinates  $(\eta,X,Y,Z)$, the light signal sent at $\eta_{A0}$ by the clock {\bf a} to the clock {\bf b} which receives the signal at $\eta _{B0}$.
After the clock {\bf b} sends back the $n \sigma$ bits of compressed information of the received time $\eta _{B0}$, the clock {\bf a} gets a complete set of the time data at the time $\eta'_{A1}=\eta _{A1}+\tau \sigma$, where $\eta _{A1}$ is the true received time of the first pulse from the clock {\bf b} and $\tau :=\tau _0 n$.
Among the times $\eta_{A0}$, $\eta _{B0}$ and $\eta' _{A1}$, we can impose the synchronization condition as in the original Einstein setting~\cite{Einstein1905},
\bea
	&&\eta _{B0}=\fr{1}{2} (\eta _{A0}+\eta _{A1})=\fr{1}{2} (\eta _{A0}+\eta' _{A1}-\tau \sigma).
\eea   

 We assume that the clocks everywhere in the frame $(\eta,X,Y,Z)$ are synchronized and adopt the reading of the clock and measure nearby an observer sitting in the rest frame as the time and space coordinates  $(t_{obs},x_{obs},y_{obs},z_{obs})$ even if the observer is moving relative to the clocks and measures.} Following the original paper by Einstein~\cite{Einstein1905} as reproduced in Appendix A, we see that the true received time $\eta$ as well as the spatial coordinates $X,Y,Z $ of the clock {\bf a} in the moving frame is related by the Lorentz transformation
 to the rest frame coordintes $(t,x,y,z)$ as 
 \bea
 &&\eta  =\fr{t-\fr{v}{c^2}x}{\sqrt{1-\left (\fr{v}{c}\right )^2}}, \label{eq:etaz}\\
 &&X=\fr{x-vt}{\sqrt{1-\left ( \fr{v}{c} \right )^2}}, \label{eq:Xz}\\
 &&Y=y ,\\
 &&Z=z.
\eea 

According to the previous discussion the apparent received time $\eta_{obs}:=\eta'$ is related to the true received time $\eta _{\rm true}$ by
$\eta_{obs}=\eta_{\rm true}+\tau\sigma$. We see that
 \bea
	\eta_{obs}
		&=&\fr{t_{\rm true}-\fr{v}{c^2}x_{\rm true}}{\sqrt{1-\left (\fr{v}{c}\right )^2}}+\tau\sigma \nonumber \\
		&=&\fr{\left ( t_{\rm true}+\fr{\tau \sigma}{\sqrt{1-\left (\fr{v}{c} \right )^2}}\right )-\fr{v}{c^2}\left ( x_{\rm true}+\fr{\tau\sigma v}{\sqrt{1-\left (\fr{v}{c} \right )^2}}\right ) }{\sqrt{1-\left (\fr{v}{c} \right )^2}}\nonumber \\
		&=&\fr{(t_{\rm true}+u^0 \tau \sigma )-\fr{v}{c^2}(x_{\rm true}+u^1 \tau \sigma ) }{\sqrt{1-\left (\fr{v}{c} \right )^2}}\;,\label{t-u}
\eea
where $u^0 :=\fr{1}{\sqrt{1-\left (\fr{v}{c} \right )^2}}$, $u^1:=\fr{v}{\sqrt{1-\left (\fr{v}{c} \right )^2}}$ and $x_{\rm true}$ is the true recieved coordinate. Similarly, from \eq{eq:Xz} $X_{obs}$ can be rewritten in terms of $t,x,u^0,u^1,\tau\sigma$ as
\bea
 X_{obs}=X=\fr{( x_{\rm true}+u^1 \tau \sigma )-v(t_{\rm true}+u^0\tau \sigma)}{ \sqrt{1-\left (\fr{v}{c} \right )^2}}\label{X}\;.
\eea
Writing  $t_{obs}=t+u^0\tau \sigma,x_{obs}= x+u^1 \tau \sigma$, we see that the Lorentz transformation holds also for  the apparently received time and coordinate,
\bea
\eta_{obs}=\fr{t_{obs}-\frac{v}{c^2}x_{obs}}{ \sqrt{1-\left (\fr{v}{c} \right )^2}}\;,\label{t_app}\\
X_{obs}=\fr{x_{obs}-vt_{obs}}{ \sqrt{1-\left (\fr{v}{c} \right )^2}}\;.\label{app_X}
\eea

The 4-vector $(u^0,u^1,0,0)$  in Eqs.~(\ref{t-u}) and (\ref{X}) can be generalized to $u^\mu$ by rotating spatial components of the 4-vector $u^{\mu}$. 
Now Eqs.~(\ref{t_app}) and (\ref{app_X}) can be summarized as
\bea
	X^\mu_{obs} =\Lambda ^\mu _{~\nu} x^\nu_{obs}\;,
\label{LT}
\eea
for 
\bea
 x^\nu_{obs}:= x^\nu+ u^\nu\tau \sigma\;,
\eea
where $\Lambda ^\mu _{~\nu}$ is the Lorentz transformation matrix.

  In the present work we take the apparent coordinates $(t_{obs},x_{obs})$, which correspond to the readings of the clock and measure, as the coordinate to specify the space-time position, e.g., as the argument of fields.
  However, the metric should be primarily defined by the true coordinates $x^\nu$ as
\bea
 ds^2_{1}=\eta_{\mu\nu}dx^{\mu} dx^{\nu}\;,\label{ds_true}
\eea
where $(\eta_{\mu\nu})=Diag(-1,1,1,1)$ is the Minkowski metric tensor, because the light travels along the null line $ds^2_{1}=0.$  
  In terms of the observed coordinates $x^\mu_{obs}$ the metric is expressed as
  
  \bea
 ds^2_{1}=\eta_{\mu\nu} (dx^\mu_{obs}- u^\mu\tau d\sigma)(dx^\nu_{obs}- u^\nu\tau d\sigma) \;.\label{clkdss}
\eea
At this stage we are convinced that the introduction of the velocity vector $u^\mu$ is necessary if we express the metric in terms of the observed coordinates in which the time delay of communication is taken into account.
We can consistently describe all the introduced fields including the velocity as  functions of the observed coordinates $x^\mu_{obs}$ thanks to the Lorentz transformation property (\ref{LT}). 
The notion of the observed coordinates becomes crucial to interpret physical consequences of the theory which contains the time-like vector field as well as the gravitational field as dynamical variables as we shall remark in VIA.
For notational simplicity hereafter we omit the suffix ``obs" from $x^\mu_{obs}$  and simply write it as $x^\mu$ without much confusion.

  So far is the special relativity with the minimum time duration for the communication taken into account. We now turn to the general relativistic extension~\cite{EinGR}.
 From \eq{clkdss}, it is natural to generalize the metric as
 \bea
	ds_1^2=g_{\mu \nu }(x)(dx^\mu -u^\mu (x)\tau d\sigma )(dx^\nu -u^\nu (x)\tau d\sigma )\;, \label{metric_prot}
 \eea
 where $g_{\mu\nu}(x) $ is the metric tensor field and $u^\mu (x)$ is a 4-vector field of the two close points
A and B located in the neighborhood of $x$. For the 4-vector field $u^\mu(x)$ let us assume
 \bea
	g_{\mu \nu}(x) u^\mu(x) u^\nu (x) =-1\;. \label{clkNorm}
 \eea
Note that in the geometrical picture the operational procedure of the light signal exchanges is implicitly used to establish the expression of the line element.

By adopting the equivalence principle, we can locally choose that $g_{\mu \nu}(x)|_{x=x_0}=\eta_{\mu\nu}$ (the \emph{Minkowski} metric)
   and $u^{\mu}(x)|_{x=x_0}=(1,0,0,0)$ (a clock co-moving with an inertial frame), we reproduce the flat metric \eq{clkdss}\footnote{The velocity field $u^{\mu}(x)$ introduced here has the 	
	operational meaning
   that $u^{\mu}(X(\eta))=\frac{dX^{\mu}(\eta )}{d\eta}$ for the true 
	proper time $\eta$, that is the 4-velocity $\fr{dX^\mu(\eta)}{d\eta}$ of the clock moving along the space-time trajectory $X^\mu(\eta)$ is 	
	identical to a time-like vector field at the space-time point $X(\eta)$ 	for the hypothetical clock. Further we assume that \eq{clkNorm} holds
	even if the argument $x$ of the 4-vector field $u^\mu(x)$ is not equal 	
	to the trajectory $X^\mu(\eta )$ of the clock. Here we remind the reader that the coordinates $x^{\mu}$ are operationally defined by the 
	readings of the
	hypothetical measure and clock there.}.
  
 Instead of \eq{metric_prot}, from the reason described  below it is better to consider the total metric as
 \bea
	ds^2=g_{\mu\nu}(x)(dx^\mu -u^\mu \tau d\sigma )(dx^\nu -u^\nu \tau d\sigma )+(a+1) \tau ^2(d\sigma )^2\;,
 \eea
 where the metric in the information space is added to the original metric by the Pythagorean rule, and $a+1$ is a positive constant to be determined by experiment.
 \begin{figure}[h]
  \begin{center}
    \includegraphics[width=0.5\columnwidth ,clip]{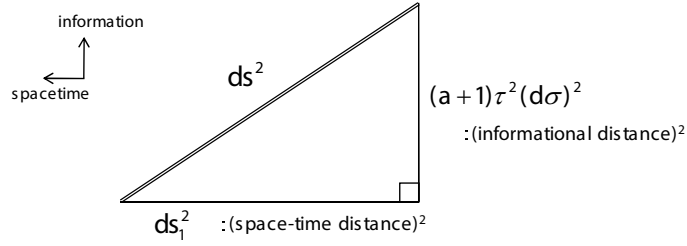}
  \end{center}
  \caption[Extension of space-time to the information space direction]{Extension of space-time to the information space direction\newline Pythagorean construction of the metric for the space-time and information space}
  \label{fig:STwithInfo2}
\end{figure}
Hereafter, the discussion will be based on the space-time-information metric,
\bea
	ds^2=g_{\mu \nu}dx^\mu dx^\nu -2\tau u_\mu dx^\mu d\sigma +a(\tau d\sigma )^2 \ =:g_{\hat{\mu} \hat{\nu}}dx^{\hat{\mu}}dx^{\hat{\nu}},
\eea
where $x^\mu$ is the ordinary space-time coordinates whose Greek indices run from 0 to 3, and $\sigma$ is the coordinate of the information space corresponding to the index 4, and the $x^{\hat{\mu}}$ collectively indicates the ordinary space-time coordinates and information space whose hatted Greek index runs from 0 to 4.
%
\section{CLASSIC  tests \label{sec:classic}}
We have put the information coordinate $\sigma$ on the same footing of the space-time coordinates. One of the merits is to quickly see the equation of motion for $\sigma$ is that the second derivative of $\sigma$ with respect to the proper time vanishes and therefore the $\sigma$ is linear in the proper time so that it
is consistent with the assumption that the proper time delay is proportional to the bits of the information.

\subsection{Geodesic equation}\label{sec:geo}
%
Starting with the action for a point particle in the 5 dimensional space-time,
\bea
	S=\int d\eta \fr{1}{2} g_{\hat{\mu}\hat{\nu}}(X^\alpha (\eta )) 
		\fr{dX^{\hat{\mu}}}{d\eta}\fr{dX^{\hat{\nu}}}{d\eta}\;.
\eea
where $X^\alpha (\eta)$ is the world line of the point particle, 
we have the geodesic equation given by
  \bea
\frac{d^2X^{\hat\lambda}}{d\eta^2}+\hat{\Gamma}^{\hat{\lambda}}_{\hat\mu\hat\nu}\frac{dX^{\hat\mu}}{d\eta}\frac{dX^{\hat\nu}}{d\eta}=0 \;,\label{5geo}
 \eea
where $\eta$ is the true proper time.\\
\noindent 
Decompose the space-time and informational components in the above equation and use the list of the Christoffel symbols (\ref{Chr-lmn}) through (\ref{Chr-444}) to see that
 \bea
&\frac{d^2X^{\lambda}}{d\eta^2}+{\Gamma}^{\lambda}_{\mu\nu}\frac{dX^{\mu}}{d\eta}\frac{dX^{\nu}}{d\eta}+
2{\Gamma}^{\lambda}_{4\nu}\frac{d\sigma}{d\eta}\frac{dX^{\nu}}{d\eta}+
{\Gamma}^{\lambda}_{44}(\frac{d\sigma}{d\eta})^2 \nonumber \\
&=\frac{d^2X^{\lambda}}{d\eta^2}+{\Gamma}^{\lambda}_{\mu\nu}\frac{dX^{\mu}}{d\eta}\frac{dX^{\nu}}{d\eta}-\tau u^{\nu}D_{\nu}u^{\lambda}\frac{d\sigma}{d\eta} \nonumber \\
&=(1-\tau \frac{d\sigma}{d\eta})[\frac{d^2X^{\lambda}}{d\eta^2}+{\Gamma}^{\lambda}_{\mu\nu}\frac{dX^{\mu}}{d\eta}\frac{dX^{\nu}}{d\eta}]=0\;,
 \eea
where $\fr{dX^{\hat{\mu}}}{d\eta}=u^\mu(X(\eta))$ is used in the second line.
 This reproduces the standard geodesic equation in 4-dimensions unless the prefactor $1-\tau \frac{d\sigma}{d\eta}$ vanishes. Moreover, the two of the classic three tests of general relativity, the bending
 of light by the Sun and the perihelion advance of Mercury~\cite{EinMer},  remain intact provided that the given gravitation field is close to the Schwartzschild metric in the scale of the solar system, which will
be justified  by the spherically symmetric solution obtained in V.
 Only the redshift formula will be modified as we shall discuss in the next subsection.
 
 For $\hat\lambda=4$, the 5-dimensional geodesic equation (\ref{5geo}) simply reduces to
  \bea
\frac{d^2\sigma}{d\eta^2}=0,
 \eea
which means that the coordinate $\sigma$ in the information space is linear in the proper time $\eta$. The implication of the essential identification of the communicated information amount $\sigma$ with
the proper time $\eta$ is suggestive from the informational view of time.
One might imagine that the $\sigma$ is the length of a history book paged by the proper time $\eta$. We see the precise relation between the two as
 \bea
g_{\hat\mu\hat\nu}\frac{dX^{\hat\mu}}{d\eta}\frac{dX^{\hat\nu}}{d\eta}=-1+2\tau b +a\tau^2 b^2=const\;,
\eea
where $\sigma=b\eta$, assuming that $\tau$, $\sigma$ and $a$ are constant. Note that the requirement that $\frac{dX^{\hat\mu}}{d\eta}$ be time-like
ensures that the prefactor $1-\tau \frac{d\sigma}{d\eta}$ does not vanish so that
  the claim that the standard geodesic equation in 4-dimensions is reproduced is {\it a posteriori} justified. 
  
\subsection{Red shift}\label{sec:reds}
The proper time $d\eta$ given by
\bea
	c^2d\eta^2=-g_{\mu\nu}(dx^\mu -\tau u^\mu d\sigma )(dx^\nu -\tau u^\nu d\sigma )-(a+1)\tau^2 d\sigma ^2
\eea
 becomes for the comoving clock $u^{i}=0, u^{0}=\frac{1}{\sqrt{-g_{00}}(x)}:=\frac{1}{f(x)}$ and for small $\tau$ 
 \bea
	c^2d\eta^2\approx -(f(x))^2[(dx^0)^2-\frac{2\tau}{f(x)}d\sigma dx^0]+O(\tau ^2).
\eea
Then we see that 
 \bea
	c d\eta\approx f(x)dx^{0}[1-\frac{\tau}{f(x)}\frac{d\sigma}{dx^0}]+O(\tau^2).
\eea

Let the proper time durations at the position $x_A$ and  $x_B$ be $d\eta_A$ and $d\eta_B$, which are communicated by a light signal at the coordinate time $x^0$ is related by
 \bea
	\frac{d\eta_A}{d\eta_B}
	\approx \frac{f(x_A)}{f(x_B)}\frac{1-\frac{\tau}{f(x_A)}\frac{d\sigma}{dx^0}}{1-\frac{\tau}{f(x_B)}\frac{d\sigma}{dx^0}}.
\eea
For the weak potential $\phi(x)$, using $f(x)\approx1+\frac{\phi(x)}{c^2}$, we arrive at the red shift formula

 \bea
 \fr{\omega _B}{\omega _A}-1=
 \frac{d\eta_A}{d\eta_B}-1
 \approx \frac{\phi(x_A)-\phi(x_B)}{c^2}[1+\tau\frac{d\sigma}{dx^0}].
\eea
The last factor exhibits the deviation from the standard red shift  formula~\cite{EinGRed}.
This deviation can be experimentally tested in principle and the minimum time $\tau$ may be evaluated. Perhaps more dramatically, consider the vertical trip of up and down of light between two positions of different heights. The standard red shift formula tells us that
the red shift and the blue shift exactly cancel. In the present case, however, they are not canceled away but the factor $1+\tau\frac{d\sigma}{dx^0}$ multicatively accumulates.
\section{Symmetry and Action}\label{sec:symmAction}

So far is a heuristic and operational introduction of the unit time-like vector $u^\mu(z)$. One of the demerits of the operational construction is that the whole picture is not easy to grasp, while each step is unambiguous. Einstein apparently switched to the deductive approach to establish the dynamics of the gravitational field in his 1915 paper \cite{EinGR}. In our present work we follow his path by starting with the gauge symmetry of the informational metric. We then impose the general coordinate invariance to the action in 5-dimensional informational space-time, which is a straightforward extension of the above gauge invariance.


 We can associate an arbitrary amount of information  $\sigma$ at each event. This gauge freedom induces
the gauge transformation for the velocity and the gravitational fields. 

\subsection{Symmetry}\label{sec:symm}

The information coordinate $\sigma$ can in general be set up differently at each space-time point $x$, although we took it homogeneous for simplicity in the previous section.

\noindent Now consider the infinitesimal transformation $\sigma\rightarrow \sigma+\phi(x)$ which makes the $\sigma$ inhomogeneous.
The metric becomes
\bea
ds^2&=&g_{\mu \nu}dx^\mu dx^\nu -2\tau dx^\mu u_\mu d\sigma +a\tau^2(d\sigma )^2 \nonumber \\
&\rightarrow& g_{\mu \nu}dx^\mu dx^\nu -2\tau dx^\mu u_\mu d\sigma +a\tau^2(d\sigma )^2 \nonumber\\
&&-\tau g_{\mu \nu}(dx^\mu u^\nu+dx^\nu u^\mu)\partial_\alpha \phi dx^\alpha+2a\tau^2\partial_\alpha \phi dx^\alpha d\sigma +O(\phi^2).
\eea 
The line element can be made invariant by the infinitesimal gauge transformation
\bea
\label{u}
&&u_\mu\rightarrow u_\mu+a\tau \partial_\mu \phi, \label{eq:gauge_u}\\
&&g_{\mu \nu}\rightarrow g_{\mu \nu}+\tau(u_\mu\partial_\nu \phi+u_\nu\partial_\mu\phi).
\label{eq:gauge_g}
\eea
It is a key observation that the gauge transformation can be embedded in the 5-dimensional coordinate transformation
 $\hat \delta g_{\hat\mu\hat\nu}= -{\cal L}_\xi g_{\hat\mu\hat\nu}= -\uf \nabla _{\hat\mu}\xi_{\hat\nu}-\uf \nabla _{\hat\nu}\xi_{\hat\mu}$ for
the infinitesimal general coordinate transformation $x^{\hat\mu}\rightarrow x^{\hat\mu}+\xi^{\hat\mu}$. 
Restricting $\xi^{\hat\mu}$ as $\xi^\mu=0\;,\xi^4=\phi(x)$ the coordinate transformation induces the change\footnote{Note that under the normalization $u^\alpha u_\alpha =-1$, $u_\alpha \hat{g}^{\alpha \mu}=u_\alpha (g^{\alpha \mu}+\fr{1}{a+1}u^\alpha u^\mu)=u^\mu-\fr{1}{a+1}u^\mu=\fr{a}{a+1}u^\mu$. By this relation two connection terms $g_{4\alpha} \Gamma ^\alpha _{\mu 4} \xi ^4$ and $g_{44}\Gamma^4_{\mu 4} \xi ^4$ appeared in $\delta g_{\mu 4}$ cancel out and then we obtain $\delta g_{\mu 4}=-\uf \nabla _\mu \xi _4-\uf \nabla_4 \xi _\mu =\uf \nabla_\mu \xi _4=-a\tau ^2 \pa _\mu \phi$.} $\delta g_{\mu 4}=-a\tau^2\partial_{\mu}\phi(x)$ and therefore  the gauge transformation $\delta u_{\mu}=a\tau\partial_{\mu}\phi(x)$
   for the vector field $u_{\mu}$. The 4-dimensional metric transforms as $\delta g_{\mu\nu}=\tau(u_\mu\partial_\nu\phi+u_{\nu}\partial_{\mu}\phi)$.
   
In what follows we are going to introduce the dynamics of the metric $g_{\mu\nu}$ and the velocity vector $u^\mu$ on the basis of invariance principle. The informational gauge transformation stated above together with the conventional 4-dimensional
general coordinate transformation are embedded in the general coordinate transformation in 5-dimensions. For the action it is natural to impose the invariance under the 5-dimensional coordinate transformation. Further we need the Einstein gravity in it to reproduce
the overwhelming success of general relativity. Then the simplest choice at the moment would be the Einstein-Hilbert action in 5-dimensions. As we shall see by the (4+1)-decomposition in the next section, the action contains the Einstein-Hilbert action for the metric tensor and the Maxwell-like action for the vector field as well as the Lagrange multiplier term, which also plays the role of cosmological energy source.


\subsection{Action\;\;- gravity part-}\label{sec:action}

We are now going to derive the 4-dimensional effective action on the basis of the gauge symmetry. For that purpose it is convenient to start from the 5-dimensional Einstein-Hilbert action assuming the (4+1) decomposition of the metric:
\bea
	ds^2=g_{\hat{\mu} \hat{\nu}}dx^{\hat{\mu}}dx^{\hat{\nu}}
	&=&g_{\mu\nu} (x)(dx^\mu -u^\mu  \tau d\sigma )(dx^\nu -u^\nu \tau d\sigma ) +(a+1)\tau^2 d\sigma ^2 \;, \label{metric}
\eea
where the hatted Greek indices, $\hat{\mu}$ run from 0 to 4, the unhatted Greek indices, $\mu$ run from 0 to 3, and the 4 indicates the coordinate $\sigma$ of the information space. The vector field $u^\mu (x)$ introduced in the previous section \ref{sec:intro} physically means the direction of time and depends only on the space-time coordinate $x^\mu$ by assumption.

The parameter $a$ denotes the scale of information space. Note that in the definition of synchronization in the previous section \ref{sec:intro}, the velocity $u^\mu (x)$ of the ubiquitous sender and reciever at $x$ is introduced. We assume that the space-time metric $g_{\mu\nu}(x)$ has no dependence on information coordinate $\sigma$  and the homogeneity of the scale $a$ of the information space. 

 Now we introduce the 5-dimensional Einstein-Hilbert action $S_0$ and the additional one $S_1$ for the normalization constraint to $u^\mu$,
\bea
	&&S_0 =\fr{c^4}{16\pi G}\int d^4x d\sigma \sqrt{-\det(\hat{g}_{\hat{\mu}\hat{\nu}})}\hat{R}\;,\label{Szero}  \\
	&&S_1 =\int d^4x d\sigma \sqrt{-\det(\hat{g}_{\hat{\mu}\hat{\nu}})}\lambda (g_{\mu\nu}u^\mu u^\nu +1)\;,\label{Sone}
\eea 
where $\hat{g}_{\hat{\mu}\hat{\nu}}$ is the 5-dimensional metric component, $\hat{R}$ is the 5-dimensional Ricci scalar and the $\lambda$ is the Lagrange multiplier which makes $u^\mu$ a unit time-like vector field.
\footnote{When calculating the metric and the Ricci curvature, we should not impose constraint condition before applying variational method, $g_{\mu\nu}u^\mu u^\nu=-1$ on them and therefore maintain the derivative terms of $u_\mu u^\mu$. But as a result of the partial integration, such derivatives and the terms including $u_\mu u^\mu$ are reduced to the terms proportional to $(u_\mu u^\mu +1)$, that is, $f(x^\mu)(u_\mu u^\mu +1)$. This means such terms can be absorbed into the Lagrange multiplier by redefining it in the action \eq{Sone} such as $\lambda '=\lambda + f(x^\mu )$. In this sense, starting from \eq{imet2} is justified and we effectively do not need to differentiate $u_\mu u^\mu$ included in 5-dimensional determinant $\sqrt{-\det g_{\hat{\mu}\hat{\nu}}}=\sqrt{-(a-u_\alpha u^\alpha ) \det{g_{\mu\nu}}}$.}

 Before applying the variational method to the total action $S=S_0+S_1$, we shall decompose the metric (\ref{metric}) and rewrite  $\hat{R}$ in terms of the 4-dimensional quantities. They are the 4-dimensional  scalar curvature and the square of the anti-symmetric tensor, which comes from the off-diagonal components of the Christoffel symbol
\bea
	\uf \Gamma ^\lambda _{4\mu}=-\fr{1}{2}\tau \hat{g}^{\lambda \rho}A_{\mu \rho}\;,
\eea  
where $A_{\mu \rho}:=D_\mu u_\rho -D_\rho u_\mu$, $D_\mu$ is the 4-dimensional covariant derivative satisfying $D_\lambda g_{\mu\nu}=0$, and the inverse metric $\hat{g}^{\mu\nu}:=\hat{g}^{\mu\nu}=g^{\mu\nu}+\fr{u^\mu u^\nu}{a+1}$. The action \eq{Szero} is superficially similar to the Kaluza-Klein action. However, in our case this is just a convenience to obtain invariant action under the gauge transformation \eq{eq:gauge_u} and \eq{eq:gauge_g} which is embedded in the 5-dimensional coordinate transformation as mentioned before. More precisely note that this inverse metric looks similar to the one in the Kaluza-Klein theory~\cite{KK} but not quite. The physical electromagnetic vector field in the Kaluza-Klein theory  is defined in the lower indices, while  in our set-up the velocity field $u^\mu$ is defined in the upper indices. 
Furthermore the action contains the energy density term in the constraint (\eq{Sone}).    \\

Using the Ricci curvature component in Appendix, \eq{tensor:R44}, \eq{tensor:R4m} and \eq{tensor:Rmn}, we obtain the 5-dimensional Ricci scalar curvature in the 4+1 decomposed form,
\bea
	\hat{R}
		&=&\hat{g}^{\hat{\alpha}\hat{\beta}}R_{\hat{\alpha}\hat{\beta}}
		 =\hat{g}^{\alpha\beta}R_{\alpha\beta}+\hat{g}^{4\beta}\uf R_{4\beta}+\hat{g}^{\alpha 4}R_{\alpha 4}+\hat{g}^{44}R_{44} \nonumber \\
		&=&\hat{g}^{\alpha\beta}R_{\alpha\beta} 
		+\fr{1}{4}\fr{1}{a+1}\hat{g}^{\alpha \rho}\hat{g}^{\beta \sigma}A_{\alpha \rho}A_{\beta\sigma} 
		-\fr{1}{a+1}u^\alpha D_\lambda (\hat{g}^{\lambda\rho}A_{\alpha \rho} ) \nonumber \\ 
		&&+\fr{1}{a+1}\hat{g}^{\alpha \beta} 
			\Bigl [-D_\rho (u^\rho D_\alpha u_\beta )+D_\beta (u^\rho D_\alpha u_\rho )-\fr{u^\rho u^\sigma}{a+1} (D_\alpha u_\rho )D_\beta u_\sigma 
		+\fr{1}{2} \bigl \{ (D_\lambda u_\alpha )D^\lambda u_\beta -(D_\alpha u_\lambda )D_\beta u^\lambda \bigr \} \Bigr ]\;. \label{RicSc}
\eea  

By the explicit determinant expansion method as shown in Appendix \ref{app:determinant}, we obtain
\bea
	\det (\hat{g})=\tau^2 (a-g_{\xi \eta}u^\xi u^\eta)  \det (g_{\alpha \beta} ).
\eea 

Then the total action $S$ becomes,
\bea
	S	&=&\fr{c^4}{16\pi G}\int 
		\sqrt{-(a-g_{\xi \eta}u^\xi u^\eta)  \det (g_{\alpha \beta} )}d^4x\tau d\sigma \Bigl [\hat{g}^{\alpha\beta}R_{\alpha\beta} 
		+\fr{1}{4}\fr{1}{a+1}\hat{g}^{\alpha \rho}\hat{g}^{\beta \sigma}A_{\alpha \rho}A_{\beta\sigma}
		-\fr{1}{a+1}u^\alpha D_\lambda (\hat{g}^{\lambda\rho}A_{\alpha \rho} ) \nonumber \\
		&&+\fr{1}{a+1}\hat{g}^{\alpha \beta} 
			\Bigl \{ -D_\rho (u^\rho D_\alpha u_\beta )+D_\beta (u^\rho D_\alpha u_\rho )-\fr{u^\rho u^\sigma}{a+1} (D_\alpha u_\rho )D_\beta u_\sigma 
			+\fr{1}{2} \bigl \{ (D_\lambda u_\alpha )D^\lambda u_\beta -(D_\alpha u_\lambda )D_\beta u^\lambda \bigr \} \Bigr \} \nonumber \\
		&&+\frac{16 \pi G}{c^4}\lambda (g_{\mu\nu}u^\mu u^\nu +1) \Bigr ]\;, 
		\label{Stotal01}
\eea 
where $\lambda$ is the Lagrange multiplier for the normalization constraint of $u^\mu$.
  To simplify \eq{Stotal01} we notice that
\bea
&& \sqrt{-(a-u^\xi u_\xi )\det (g_{\delta \gamma})}2B^\mu u^\alpha D_\mu u_\alpha \nonumber \\
	&&=\sqrt{-(a-u^\xi u_\xi )\det (g_{\delta \gamma})}B^\mu D_\mu (u^\alpha u_\alpha +1) \nonumber \\
	&&=
			\left [ D_\mu \left \{ \sqrt{-(a-u^\xi u_\xi ) \det (g_{\delta \gamma})}B^\mu (u_\alpha u^\alpha +1 ) \right \} 
			-D_\mu \left ( \sqrt{-(a-u^\xi u_\xi )\det (g_{\delta \gamma})}B^\mu\right )(u_\alpha u^\alpha +1) \right ] \;,
						\label{eq:uDu}
\eea
for  an arbitrary vector field $B^\mu$.
Note that the first term in \eq{eq:uDu} is zero with an appropriate boudary condition and $B^\mu u^\alpha D_\mu u_\alpha$ in the action becomes a form of  $f(x^\mu) (u_\alpha u^\alpha +1)$ upon partial integration, which has the same form as the constraint term and therefore can be absorbed into the definition of the Lagrange multiplier $\lambda$. 
   Manipulating the third term in \eq{Stotal01} as in Appendix \ref{app:simplification}, we obtain 
\bea
	-\fr{1}{a+1}u^\alpha D_\lambda (\hat{g}^{\lambda \rho}A_{\alpha \rho})
	&=&-\fr{\hat{g}^{\alpha \beta} \hat{g}^{\lambda \rho}}{2(a+1)}A_{\alpha \rho}A_{\beta \lambda}\;.\label{t3rd}
\eea
Similarly the fourth term in \eq{Stotal01} using \eq{app:t4_2_eq3} becomes
  \bea
    &&\fr{1}{a+1}\hat{g}^{\alpha \beta} 
			\Bigl \{ -D_\rho (u^\rho D_\alpha u_\beta )+D_\beta (u^\rho D_\alpha u_\rho )-\fr{u^\rho u^\sigma}{a+1} (D_\alpha u_\rho )D_\beta u_\sigma 
			+\fr{1}{2} \bigl ( (D_\lambda u_\alpha )D^\lambda u_\beta -(D_\alpha u_\lambda )D_\beta u^\lambda \bigr ) \Bigr \} \nonumber \\
		&=& \fr{1}{(a+1)^2} u^\rho (D_\rho u^\alpha )u^\beta  D_\alpha u_\beta +\fr{1}{2}\fr{1}{(a+1)^2}a_\lambda a^\lambda\;,\label{t4th}
\eea
where $a_\lambda :=u^\alpha D_\alpha u_\lambda$. Substituting \eq{t3rd} and \eq{t4th} into \eq{Stotal01}, the action is further simplified to
\bea
	S&=&\fr{c^4}{16\pi G}\int \sqrt{-(a-g_{\xi \eta}u^\xi u^\eta)  \det (g_{\alpha \beta} )}d^4x\tau d\sigma \Bigl [\hat{g}^{\alpha\beta}R_{\alpha\beta} -\fr{g^{\alpha \beta}g^{\rho \sigma}}{4(a+1)}A_{\alpha \rho}A_{\beta \sigma}+\fr{1}{(a+1)^2}u^\rho (D_\rho u^\alpha )u^\beta  D_\alpha u_\beta \nonumber \\
	&&+\frac{16\pi G}{c^4}\lambda ' (g^{\alpha \beta} u_\alpha u_\beta +1)\Bigl] \label{t5th},
\eea
where \eq{app:AAaa} is used. Note that the third term in \eq{t5th} can be partially integrated to yield the term proportional to $(g^{\alpha \beta}u_\alpha u_\beta +1)$ which can be absorbed into a part of the following constraint term and thus $\lambda '$ is rewritten as $\lambda ''$ by this redefinition of the Lagrange multiplier.

We finally obtain a compact form of the action, rewriting the redefined Lagrange multiplier $\lambda''$ above as $\lambda$ for simplicity,
\bea
	S=&\fr{c^4}{16\pi G}\int \sqrt{-(a-g_{\xi \eta}u^\xi u^\eta)  \det (g_{\alpha \beta} )}d^4x\tau d\sigma \Bigl [\hat{g}^{\alpha\beta}R_{\alpha\beta} -\fr{g^{\alpha \beta}g^{\rho \sigma}}{4(a+1)}A_{\alpha \rho}A_{\beta \sigma}+\frac{16\pi G}{c^4}\lambda (g^{\alpha \beta} u_\alpha u_\beta +1)\Bigl ],
\label{action}
\eea
where $\hat{g}^{\alpha\beta}=g^{\alpha\beta}+\fr{u^\alpha u^\beta}{a+1}$, $R_{\alpha \beta}$  is the Ricci tensor and the field strength of the velocity field is  $A_{\alpha \rho}:=D_\alpha u_\rho -D_\rho u_\alpha$.
The first term represents the Einstein-Hilbert action with the slight modification by the $\hat{g}^{\alpha\beta}$, the second term is the Maxwell-like action and the last term plays the triple role of the Lagrange multiplier term
to ensure the normalization of the vector field $u_\mu$, the energy density and the gauge fixing term. Only the term $g^{\mu\nu}u_{\mu}u_{\nu}$ in the action breaks this gauge symmetry
in the subsection \ref{sec:symm}. One may notice that the action (\ref{action}) does not contain the parameter $\tau$, which would appear if he introduces source term e.g., the point mass source. The situation is
parallel to the electromagnetic theory, in which the electric charge appears when the source is introduced.

 The action \eq{action} turns out to be a particular case of the parameterized action given by B\"{ o}hmer and Harko~\cite{Boemer} and similar to the action of the so-called TeVeS  theories~\cite{BekenTeves} except that we
 have not introduced the scalar field for simplicity. The action \eq{action} is also a special case of  the Einstein-Aether theory~\cite{EAE} by Jacobson and his collaborators, the comparison of which with our theory shall be discussed later. 
\subsection{Matter field}\label{sec:ActMat}
 So far is the dynamics of the space-time geometry for the metric $g^{\alpha\beta}$ and the vector field $u_\mu$, which describes the gross structure of the universe.
 We turn to the matter fields e.g., photon, electron and quark etc. in the universe, represented by a scalar field $\psi$ coupled with the cosmological fields, $g^{\alpha\beta}$ and $u_\mu$. The matter fields will modify the space-time via the Einstein equation.
 
  Starting with the 5-dimensional action 
 \bea
 S_{{\rm matter}}=-\int d^{5}x\frac{1}{2} \sqrt{-\det( \hat{g}_{\hat{\mu} \hat{\nu}})}
 \hat{g}^{\hat{\mu} \hat{\nu}} \partial_{\hat{\mu}}\psi(x) \partial_{\hat{\nu}}\psi(x)\;, \label{eq:Sm}
 \eea
we obtain
 \bea	
S_{{\rm matter}}=- \int d^{4}xd\sigma\frac{1}{2} 
\sqrt{-\det( \hat{g}_{\hat{\mu} \hat{\nu}})}
\hat g^{\alpha \beta}\partial_{\alpha}\psi(x) \partial_{\beta}\psi(x),
 \eea 
where $\hat g^{\alpha \beta}=g^{\alpha \beta}+\frac{u^\alpha u^{\beta}}{a+1}$, assuming that $\psi(x)$ does not depend on the information coordinate $\sigma$.
More explicitly, the action $S_{matter}$ for the matter field $\psi(x)$ is given by
  \bea
 S_{\rm matter}
=- \int d^{4}x d\sigma\frac{1}{2} 
\sqrt{-\hat{g}} [g^{\alpha \beta}\partial_{\alpha}\psi(x) \partial_{\beta}\psi(x)+\frac{1}{a+1}(u^\alpha\partial_\alpha \psi)^2].
 \eea 
 The second term exhibits a new feature of our theory that the light velocity effectively changes depending on the vector field $u^\alpha(x)$.
 For the Minkowski space-time, $g_{\mu\nu}=\eta_{\mu\nu}$ and $u^{\mu}=(1,0,0,0)$, the velocity $v_{\psi}$ of the field $\psi$ is given by $v_{\psi}^2=\frac{c^2}{1-\frac{1}{a+1}}$. Note that $v_\psi$ here is the velocity measured by the observed apparent time rather than the true time. This point will be further discussed in VIA in the comparison with the Einstein-Aether theory. 
 The mass term and the curvature dependent term for $\psi$ can be straightforwardly  introduced which we will not discuss here.
\section{Field Equations and a spherically symmetric solution}\label{sphericaly sym}
\subsection{Field equations}\label{sec:fieldeq}
Now that we have obtained the explicit  action (\ref{action}), we are in a position to derive the variational equations with respect to the metric tensor $g^{\alpha \beta}$ and the velocity field $u_\mu$ on the basis of the action principle.
 Note that the variation of the quantity $g_{\xi\eta}u^\xi u^\eta$ contained in the prefactor in the action (\ref{action}) can be absorbed in the re-definition of the Lagrange multiplier $\lambda $ so that we can effectively ignore it in the variation.
 The variation of $\hat{g}^{\alpha\beta}=g^{\alpha\beta}+\fr{u^\alpha u^\beta}{a+1}$ with $u^\alpha=g^{\alpha\beta}u_\beta$ keeping the velocity field $u_\beta$ fixed becomes
 \bea
\delta_g\hat{g}^{\alpha \beta}=
\delta g^{\mu \nu}[\frac{1}{2}(\delta^{\alpha}_{\mu}\delta^{\beta}_{\nu}+\delta^{\alpha}_{\nu}\delta^{\beta}_{\mu})
	+\frac{1}{2(a+1)}
	\{
		(u_\mu\delta^\alpha_\nu+u_\nu\delta^\alpha_\mu)u^\beta
	+	 u^\alpha (u_\mu\delta^\beta_\nu+u_\nu\delta^\beta_\mu)
 \}]\label{variation1}
\eea  
and from \eq{action} the Lagrangian reads
\bea
	L=\hat{g}^{\alpha\beta}R_{\alpha\beta} -\fr{g^{\alpha \beta}g^{\rho \sigma}}{4(a+1)}A_{\alpha \rho}A_{\beta \sigma}+\frac{16\pi G}{c^4}\lambda (g^{\alpha \beta} u_\alpha u_\beta +1).
\eea
With the two remarks above in mind we obtain
 \bea
\frac{16\pi G}{c^4}\frac{\delta_g L}{\sqrt{-g}}
	&=&-\frac{16\pi G}{c^4}\frac{1}{2}Lg_{\mu \nu}\delta g^{\mu \nu}
	+\delta R_{\alpha\beta} \Bigl ( g^{\alpha \beta}+\fr{u^\alpha u^\beta}{a+1}\Bigr )
	+R_{\alpha\beta}\Bigl [ \frac{1}{2}(\delta^{\alpha}_{\mu}\delta^{\beta}_{\nu}+\delta^{\alpha}_{\nu}\delta^{\beta}_{\mu}) 
		+\frac{1}{2(a+1)}
		\bigl \{ (u_\mu\delta^\alpha_\nu +u_\nu\delta^\alpha_\mu)u^\beta \nonumber \\ 
	&&			+  u^\alpha (u_\mu\delta^\beta _\nu +u_\nu\delta^\beta _\mu )
		\bigr \} \Bigr ] \delta g^{\mu \nu}
 -\frac{1}{2(a+1)}g^{\alpha \beta}A_{\alpha\mu}A_{\beta\nu}\delta g^{\mu \nu}+\frac{16\pi G}{c^4} u_\mu u_\nu\delta g^{\mu \nu}\nonumber\\
 &=&\fr{u^\mu u^\nu}{a+1}\delta R_{\mu \nu }
		+\Bigl [ -\frac{1}{2}Lg_{\mu \nu}
			+R_{\mu \nu}
			+\frac{1}{a+1}(R_{\nu\alpha}u_\mu +R_{\mu\alpha}u_\nu )u^{\alpha}
			-\frac{1}{2(a+1)}g^{\alpha \beta}A_{\alpha\mu}A_{\beta\nu}
			+\frac{16\pi G}{c^4} u_\mu u_\nu \Bigr ]\delta g^{\mu \nu}\nonumber\\
 &=&\fr{u^\mu u^\nu }{a+1}\delta R_{\mu \nu}
			+\Bigl [ R_{\mu \nu}
			-\frac{1}{2}g_{\mu \nu}R_{\alpha \beta}\hat{g}^{\alpha\beta}
			+\frac{1}{a+1}(R_{\nu\alpha}u_\mu+R_{\mu\alpha}u_\nu)u^{\alpha}\nonumber \\
 &&			-\frac{1}{2(a+1)}(g^{\alpha \beta}A_{\alpha\mu}A_{\beta\nu}-\frac{1}{4}g_{\mu \nu}A_{\alpha\beta}A^{\alpha\beta})
			+\frac{16\pi G}{c^4} u_\mu u_\nu\Bigr ] \delta g^{\mu \nu},
\eea  
where in the second equality $g^{\alpha\beta}\delta R_{\alpha\beta}$ can be written in a total derivative form as in \eq{eq:gdR}, which results in $g^{\alpha\beta}\delta R_{\alpha\beta}=0$ in the action. On the other hand, $u^\mu u^\nu \delta R_{\mu \nu}$ remains non-zero and this contribution gives a deviation from the standard general relativity. The term $u^\mu u^\nu \delta R_{\mu \nu}$ can be partially integrated in the action to give
\bea
	u^\mu u^\nu \delta R_{\mu\nu}
		&=&-\fr{1}{2}D_\sigma D_\rho \Bigl [ \delta ^\rho _{~\nu}u_\mu u^\sigma +\delta ^\rho _{~\mu}u_\nu u^\sigma -g^{\sigma \rho}u_\mu u_\nu- u^\rho u^\sigma g_{\mu\nu} \Bigr ] \delta g^{\mu \nu}\;,
\eea
as shown in \eq{app:dl_uuR}.

Let us turn to the variation with respect to $u_\alpha$
\bea
\frac{16\pi G}{c^4}\frac{\delta_u L}{\sqrt{-g}}=\delta_u \hat{g}^{\mu\nu}R_{\mu\nu}+(\frac{1}{a+1}D_\mu A^{\mu\alpha}
+\frac{32\pi G}{c^4}\lambda u^\alpha)\delta u_{\alpha},
\eea  
where we have used the partial integration in the action integral. Noting 
\bea
\delta_u \hat{g}^{\mu\nu}=\frac{1}{a+1}[g^{\mu\alpha}u^\nu+g^{\nu\alpha}u^\mu]\delta u_{\alpha}\;,
\eea  
we see that
\bea
\frac{16\pi G}{c^4}\frac{\delta_u L}{\sqrt{-g}}=\Bigl [\frac{1}{a+1}(D_\alpha A^{\alpha \mu}+2R^\mu_{~\alpha}u^\alpha)
-\frac{32\pi G}{c^4}\lambda u^\mu \Bigr ]\delta u_{\mu}.
\eea

To summarize we have the two field equations
\bea
&&R_{\mu \nu}-\frac{1}{2}g_{\mu \nu}R_{\alpha\beta}\hat{g}^{\alpha\beta}
	+\frac{1}{a+1}(R_{\nu\alpha}u_\mu+R_{\mu\alpha}u_\nu)u^{\alpha}
	-\frac{1}{2(a+1)}(g^{\alpha \beta}A_{\alpha\mu}A_{\beta\nu}
			-\frac{1}{4}g_{\mu \nu}A_{\alpha\beta}A^{\alpha\beta})\nonumber \\
\label{fieldeq1}
&&\hspace{3pc} -\fr{1}{2(a+1)}D_\sigma D_\rho \Bigl [ \delta ^\rho _{~\nu}u_\mu u^\sigma +\delta ^\rho _{~\mu}u_\nu u^\sigma  -g^{\sigma \rho}u_\mu u_\nu- u^\rho u^\sigma g_{\mu\nu} \Bigr ] +\frac{16\pi G}{c^4}\lambda  u_\mu u_\nu=0,\label{Einstein}\\
&&\frac{1}{a+1}(D_\alpha A^{\alpha\mu}+2R^{\mu}_{~\alpha}u^\alpha)
+\frac{32\pi G}{c^4}\lambda u^\mu =0.\label{eq:u_field}
 \label{fieldeq2}
\eea  
Here the readers are reminded that $R^{\mu} _{~\alpha}$  is the Ricci tensor and the field strength of the normalized ($g^{\mu\nu}u_\mu u_\nu=-1$) velocity field $u^\alpha$ is  $A_{\alpha \rho}:=D_\alpha u_\rho -D_\rho u_\alpha$.
The first is the Einstein equation modified by the terms of the second order derivative of $u_\mu$ with the cosmic energy density $\lambda$. 
The physical meaning of the second Maxwell-like equation is clear; the vector field equation with a curvature dependent term and the potential term proportional to $\lambda$. The vector field $u^\alpha$ gives the energy-momentum tensor which
makes the space-time curved through ({\ref{Einstein}) as a back reaction. 
\subsection{Spherically symmetric solution}\label{sec:spherical}
We now turn to an application to astrophysics, and begin by writing down the metric, the Ricci tensor, the evolution equations for a spherically symmetric but possibly time-dependent system. Once we have decomposed the metric into 4+1 dimensional form, we can focus on the four dimensional part of the space-time while the other metric components appear as a 4-vector $u_\mu$ and a scalar $a$. A spherically symmetric 4-dimensional line element $ds_4^2$ in a diagonal form can be written as
\bea
	ds^2_4=-B(t,r)c^2dt^2+A(t,r)dr^2+r^2d\theta ^2 +r^2\sin ^2 \theta d\phi ^2\;. \label{SPmetric}
\eea  
For simplicity, we assume that the 4-vector field is spherically symmetric and is taken so as to co-move with local inertial frame, $u^\mu (r)=(u^t(r),0,0,0 )$. Usage of the constraint $u^\alpha u_\alpha=g_{tt}u^tu^t=-1$ determines the $t-$component as
\bea
	u^t(t,r)=\fr{1}{\sqrt{B(t,r)}}\;.
\eea

In the Einstein equation \eq{fieldeq1} we see that the third term, $\fr{1}{a+1}(R_{\alpha \mu}u_\nu +R_{\alpha \nu}u_\mu)u^\alpha$ and the fifth term, $P_{\mu\nu}:=-\fr{1}{2(a+1)}D_\sigma D_\rho \Bigl [ \delta ^\rho _{~\nu}u_\mu u^\sigma +\delta ^\rho _{~\mu}u_\nu u^\sigma-g^{\sigma \rho}u_\mu u_\nu-u^\lambda u^\rho g_{\mu\nu} \Bigr ] $ are the additional terms comparing with the field equations in the usual Einstein-Maxwell system. The components of the Ricci tensor are shown as \eq{eq:Rictt}--\eq{eq:Ricphph} in Appendix \ref{app:dR}.
Now let us see the $P_{\mu\nu}$ components. 
As shown in Appendix \eq{eq:Ptt} -- \eq{eq:Pphph}, each component of $P_{\mu\nu}$ is written as
\bea
	P_{tt}&=&-\fr{1}{2(a+1)}\Bigl [ -\fr{1}{2}\fr{B''}{A}+\fr{1}{4}\fr{A'}{A}\fr{B'}{A}+\fr{1}{4}\fr{(B')^2}{AB}+\fr{1}{2}\fr{\ddot{A}}{A}-\fr{1}{4}\fr{\dot{A}}{A}\fr{\dot{B}}{B}-\fr{1}{r}\fr{B'}{A}-\fr{1}{4}\fr{\dot{A}^2}{A^2}\Bigr ] \;,\\
	P_{tr}&=&0\;,\\
	P_{rr}&=&-\fr{1}{2(a+1)}\Bigl [ \fr{1}{2}\fr{\ddot{A}}{B}-\fr{1}{4}\fr{\dot{A}\dot{B}}{B^2}-\fr{1}{4}\fr{\dot{A}^2}{AB}-\fr{1}{2}\fr{B''}{B}+\fr{1}{4}\left ( \fr{B'}{B}\right )^2+\fr{1}{4}\fr{A'}{A}\fr{B'}{B}-\fr{1}{r}\fr{B'}{B} \Bigr ]\;, \\
	P_{\theta\theta}&=&-\fr{-r^2}{2(a+1)}\Bigl [ \fr{1}{2}\fr{\ddot{A}}{AB}-\fr{1}{4}\fr{\dot{A}\dot{B}}{AB^2}-\fr{1}{4}\fr{\dot{A}^2}{A^2B}+\fr{1}{2}\fr{B''}{AB}-\fr{1}{4}\fr{(B')^2}{AB^2}-\fr{1}{4}\fr{A'B'}{A^2B}+\fr{1}{r}\fr{B'}{AB} \Bigr ]\;, \\
	P_{\phi\phi}&=&-\fr{-r^2\sin ^2\theta}{2(a+1)}\Bigl [ \fr{1}{2}\fr{\ddot{A}}{AB}-\fr{1}{4}\fr{\dot{A}\dot{B}}{AB^2}-\fr{1}{4}\fr{\dot{A}^2}{A^2B}+\fr{1}{2}\fr{B''}{AB}-\fr{1}{4}\fr{(B')^2}{AB^2}-\fr{1}{4}\fr{A'B'}{A^2B}+\fr{1}{r}\fr{B'}{AB} \Bigr ] \;,
\eea
where $'$ and $\dot{}$ denote the partial derivative with respect to $r$ and $t$, respectively. The other components of $P_{\mu\nu}$ identically vanish,  $P_{r\theta}=P_{r\phi}=P_{\theta\phi}=P_{\theta t}=P_{\phi t}=0$. 

 As in \eq{eq:G12}
the $tr-$component of \eq{Einstein} becomes  
\bea
\dot{A}=0\;, \label{eq:dotA}
\eea
which tells us that $A$ is time independent.
Inspection of the time derivatives in the Einstein equations Eqs.~(\ref{eq:G11})--(\ref{eq:G44}) shows that all the time derivatives drop out of the Einstein equations by (\ref{eq:dotA}), which means that the Birkhoff theorem holds, that is,
both the  4-vector field and the metric must be static. 

Fortunately, the $rr$-component of the Einstein equation \eq{eq:G22} involves only $A,B$ and $B'$, then we can immediately obtain the expression for $A$,
\bea
A(r)=1+r\fr{B'}{B}+\fr{r^2}{\alpha ^2}\fr{(B')^2}{B^2}\;, \label{eq:A_B}
\eea  
where $\alpha^2 :=16(1+a)$ and $B':=\fr{dB}{dr}$.

Substitution of \eq{eq:dotA} and \eq{eq:A_B} into \eq{eq:G33} yields 
\bea
	G_{\theta\theta}
	&=&\fr{2(3+4a)r^2 B'}{\alpha^2 B^2}\Bigl [B'+\fr{r^2(B')^3}{2\alpha ^2B^2}+\fr{rB''}{2}\Bigr ]=0\;. \label{Sp_Beq1}
\eea
Setting $B=e^{F(r)}$ in \eq{Sp_Beq1}, we obtain
\bea
	\fr{1}{2}rF''+F'+\fr{r^2(F')^3}{2\alpha ^2}=0\;\label{eq:2ndF}.
\eea 
Note that $B'=0 \;(F'=0)$ is also a solution, which leads to the flat space-time $A=B=1$.
The equation (\ref{eq:2ndF}) can be converted into a simpler form in terms of $V$, putting $F'(r)=\fr{V(r)}{r}$, 
\bea
	\fr{dV}{dr}=-\fr{V}{r}\fr{V^2+\alpha^2 V+\alpha ^2}{\alpha ^2}\label{eq:dV}.
\eea

A solution $V(r)$ of this ordinary differential equation is found to be an implicit function of $r$, such as 
\bea
	&&\fr{r(V)}{C'}=\fr{|V-v_1|^\fr{1+\varepsilon }{2}}{|V|}\fr{1}{|V-v_2|^\fr{\varepsilon -1}{2}}\;,\label{eq:r_V}\\
\eea
where $C'$ is an integration constant and $\varepsilon  :=\fr{\alpha ^2}{\sqrt{\alpha ^4 -4\alpha ^2}}$, $v_1:=\fr{1}{2}(-\alpha ^2 +\sqrt{\alpha ^4-4\alpha ^2})$, $v_2:=\fr{1}{2}(-\alpha ^2 -\sqrt{\alpha ^4-4\alpha ^2})$.
Noting that the first factor $\fr{|V-v_1|^\fr{1+\varepsilon }{2}}{|V|} $ exhibits a Schwarzschild-like behavior and the second factor $ \fr{1}{|V-v_2|^\fr{\varepsilon -1}{2}} $ corresponds to an informational correction which goes to unity in the GR limit($\alpha\rightarrow \infty$ ,i.e., $\varepsilon  \rightarrow 1$).

Note that from the definition of $V$($F'(r)=\fr{V(r)}{r}$) we can integrate $F'(r)$ in the form
\bea
	F	&=&\int \fr{V(r)}{r}dr =\int \fr{V}{r}\fr{1}{\fr{dV}{dr}}dV\nonumber \\
	&=&\fr{\alpha ^2}{ v_1- v_2}\int \fr{-1}{V^2+\alpha ^2 V +\alpha ^2}dV\; \nonumber \\
   &=&\varepsilon \log \fr{|V+\fr{1}{2}(\alpha ^2 + \sqrt{\alpha ^4 -4\alpha ^2})|}{|V+\fr{1}{2}(\alpha ^2 -\sqrt{\alpha ^4 -4\alpha ^2})|}+f(t)\;,\label{eq:F_V}
\eea
where in the third equality, \eq{eq:dV} is used and $f(t)$ is an integration constant, which can depend on $t$. The  function $f(t)$ can be effectively made to unity by re-defining the time coordinate
\bea
	t'=N\int ^t e^\fr{f(t)}{2}dt\;,
\eea
where $N$ is a constant to be determined below. Therefore, we see that  $B (=g_{tt})$ can be made $\dot{B}=0$. 
Noting $B(r)=e^{F(r)}$, we can express $B$ in terms of $V$ as
\bea
	B(V)
		&=& N^{2}\left | \fr{V+\fr{1}{2}(\alpha ^2+\sqrt{\alpha ^4 -4\alpha ^2})}{V+\fr{1}{2}(\alpha ^2 -\sqrt{\alpha ^4 -4\alpha ^2})}\right |^\varepsilon
		\label{eq:B_V}
\eea 
To see the approach to general relativity for a large $\alpha$, we approximate Eqs.~(\ref{eq:r_V}) and (\ref{eq:B_V}) as
\bea
	&&\fr{r(V)}{C'}\simeq 1+\fr{1}{V}\;, \label{eq:ra_V}\\
	&&B(V)\simeq N^{2}
	\left | \fr{\alpha ^2 +\sqrt{\alpha ^4 -4\alpha ^2}}{\alpha ^2-\sqrt{\alpha ^4 -4\alpha ^2}}\right |^\varepsilon 
	\frac{1}{V+1}\label{eq:Ba_V} 
\eea
Combining Eqs.~(\ref{eq:ra_V}) and (\ref{eq:Ba_V}) we obtain a Schwarzschild-like\footnote{After imposing the normalization condition the combination of \eq{eq:ra_V} and \eq{eq:Ba_V} yields a Schwarzschild-like metric.} metric, 
$B(r) \simeq N^2\left ( \fr{(\fr{1}{2}(\alpha ^2+\sqrt{\alpha ^4 -4\alpha ^2})}{(\fr{1}{2}(\alpha ^2-\sqrt{\alpha ^4 -4\alpha ^2})}\right )^\varepsilon (1-\fr{C'}{r}) $.
 Then we can see that $B$ can be normalized to $1$ at $r\rightarrow \infty (V\rightarrow 0)$ (asymptotically Minkowskian) by choosing the constant $N$ as
\bea
N^2
	&=&\left | \fr{\alpha ^2 -\sqrt{\alpha ^4 -4\alpha ^2}}{\alpha ^2 +\sqrt{\alpha ^4 -4\alpha ^2}}\right |^\varepsilon 	 \;.
\eea
Finally (\ref{eq:B_V}) becomes,
\bea
	B(V)
	&=& \left | \fr{\alpha ^2 -\sqrt{\alpha ^4 -4\alpha ^2}}{\alpha ^2 +\sqrt{\alpha ^4 -4\alpha ^2}}\right |^\varepsilon 	
	\left | \fr{V+\fr{1}{2}(\alpha ^2 +\sqrt{\alpha ^4 -4\alpha ^2})}{V+\fr{1}{2}(\alpha ^2- \sqrt{\alpha ^4 -4\alpha ^2})}\right |^\varepsilon \;.\label{eq:Breg_V} 
\eea

Now let us turn to the expression for $A(V)$. From \eq{eq:Breg_V}, we see that
\bea
	&& \fr{\fr{dB(V)}{dr}}{B}=\fr{V}{r}\;, \label{eq:dB_cp}
\eea
where \eq{eq:dV} is used. From Eq.(\ref{eq:A_B}) and Eq.~(\ref{eq:dB_cp}) we obtain the expression for $A (=g_{rr})$ as a function of $V$,
\bea
	A(V)=1+V+\fr{V^2}{\alpha ^2}\;. \label{eq:A_V}
\eea   
Note that  $A\rightarrow 1+V$ in the GR limit($\alpha \rightarrow \infty$).

Finally let us see the energy density $\lambda$ (the Lagrange multiplier), which appears on the right hand side of the Einstein equation. With \eq{eq:dotA}, $G^{\rm eff}_{tt}=0$ can be reduced to the expression for $\lambda$ as a function of $V$,
\bea
	\lambda (V)=\fr{c^4}{8\pi G A(V)}\fr{V^2}{r^2\alpha ^2} \;, \label{eq:lam_V}
\eea
where the detailed derivation is shown in \eq{eq:l_V} of Appendix \ref{app:sph_sym_sol}. Note that the energy density $\lambda$ is positive definite as far as $A>0$.

Although every component of the Einstein equation is used, the field equation \eq{eq:u_field} in the spherically symmetric case has not been checked yet. 
As in Appendix \ref{app:check_u} and the  $\theta-$ and $\phi-$ components of the field equations of $u^\mu$ are identically zero and with the solution of the Einstein equation $\dot{B}=0$ and \eq{eq:dotA}, the left hand side of the $r-$component of the field equation \eq{eq:ueom_r3} is automatically satisfied. From \eq{eq:ueom_t} and \eq{eq:dotA} the left hand side of the time component of \eq{eq:u_field} reduces to
\bea
	-\fr{B'}{rB}+\fr{A'B'}{4AB}-\fr{B''}{2B}+\frac{2\pi G}{c^4} \alpha ^2 \lambda A\;.\label{eq:sph_sym_u_field}
\eea
Substituting the expressions \eq{eq:lam_V}, \eq{eq:dB_cp}, \eq{eq:A_B} and the following two relations
\bea
	&&\fr{\fr{d^2B}{dr^2}}{B}=-\fr{2V\alpha^2 +V^3}{r^2 \alpha ^2}\;,\\
	&&\fr{\fr{dA}{dr}}{A}=-\fr{V(\alpha^2 +2V)}{r \alpha ^2}\;,
\eea
where detailed derivation is shown in Eqs.~(\ref{eq:ddB}) and (\ref{eq:dA}),  we see that the quantity (\ref{eq:sph_sym_u_field}) identically vanishes, which ensures that the obtained set of expressions \eq{eq:A_B} and \eq{eq:B_V} is actually a solution of the whole set of the field equations. 

To summarize, Eqs.~(\ref{eq:A_B}),(\ref{eq:r_V}),(\ref{eq:B_V}), and (\ref{eq:lam_V}) are the solution for $V>0$,

\bea
	&&\fr{r(V)}{C'}=\fr{|V+\fr{1}{2}(\alpha ^2 - \sqrt{\alpha ^4 -4\alpha ^2})|^\fr{1+\varepsilon }{2}}{|V|}\fr{1}{|V+\fr{1}{2}(\alpha ^2 +\sqrt{\alpha ^4 -4\alpha ^2})|^\fr{\varepsilon -1}{2}},\\
	\nonumber
	&&A(V)=1+V+\fr{V^2}{\alpha ^2}\;,\\
	\nonumber
	&&B(V)=\left | \fr{\alpha ^2  -\sqrt{\alpha ^4 -4\alpha ^2}}{\alpha ^2 + \sqrt{\alpha ^4 -4\alpha ^2}}\right |^\varepsilon 
	\nonumber	
	\left | \fr{V+\fr{1}{2}(\alpha ^2 + \sqrt{\alpha ^4 -4\alpha ^2})}{V+\fr{1}{2}(\alpha ^2  - \sqrt{\alpha ^4 -4\alpha ^2})}\right |^\varepsilon , \\
	&&\lambda (V)=\fr{c^4}{8\pi G A(V)}\fr{V^2}{r^2\alpha ^2} .
\eea

Eling and Jacobson \cite{EAE_sp} also found a spherically symmetric solution with a single parameter 
($C_1$), assuming that the geometry is static. We emphasize that in our case the Birkhoff theorem is a consequence of a spherical symmetry without assuming the staticity. Their result seemingly coincides with our result.

\subsubsection{Numerical analysis}
  By taking $V$ as a parameter, our analytical solutions can be numerically plotted as a pair of two functions with the common parameter $V$ such as $\{ r(V),B(V) \}$. We will show the behavior of  $B(V), A(V), \lambda (V)$ as functions of the radius $r$ in Fig.5. Note that $\alpha =2$ corresponds to a maximal deviation from the Schwarzschild solution and $\alpha \longrightarrow \infty $ corresponds to the standard general relativistic (Schwarzschild) limit. We see that for $V<0$ the solution turns out to be physically meaningless, since $B(r)$ and $A(r)$ would  asymptotically go to zero as $r\rightarrow \infty$. Therefore, the numerical plot of the solutions is restricted to the positive parameter $V$ region and the integration constant $C'$ in \eq{eq:r_V} is set to $C'=1$ in \fig{fig:ABl}.
\begin{figure}[h]
\begin{center}
\includegraphics[width=12cm,clip]{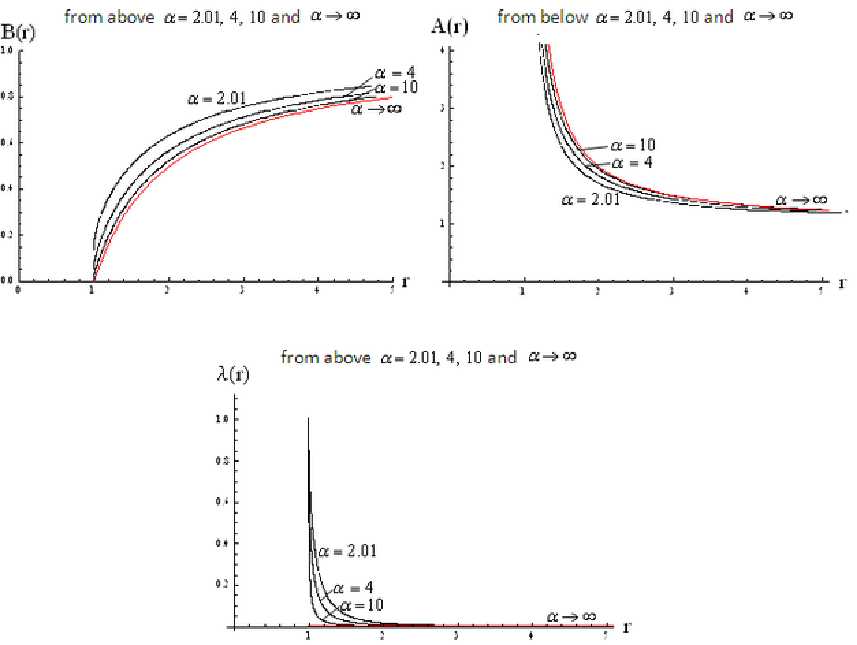}
\end{center}
\caption[The spherically symmetric solution]{The spherically symmetric solution.\newline Each graph corresponds to the metric of the $tt-$, the $rr-$ component and the energy density, $B(r)$, $A(r)$ and $\lambda (r)$, respectively. Each variable is shown in the cases of $\alpha =2.01,4,10$ and $\infty$.The $\alpha=2.01$ curve shows an almost maximally deviated curve from the Schwarzschild solution. On the other hand, the curves with $\alpha =10$ and $\alpha =\infty$(GR-limit) are close to each other. We can see that at remote distance $B(r),A(r)$ and $\lambda(r)$ behave almost like the Schwarzschild solution even at $\alpha =2.01$, but  near $r=1$ in $\alpha =2.01$ the behavior of metric deviates from the Schwarzschild case.}
\label{fig:ABl}
\end{figure} 

In \fig{fig:ABl}, the metric components $A(r)$ and $B(r)$, and the energy density $\lambda (r)$ whose dynamics is determined by $tt-$component of the effective Einstein equation, approach the Schwarzschild solution when $r\longrightarrow \infty$ for a fixed $\alpha$ or $\alpha \longrightarrow \infty$ for a fixed radius $r$. When $\alpha$ is small, the informational effect to gravity becomes more apparent and our solution significantly deviates from the Schwarzschild solution near $r=1$ in some unit. 
Note that from \eq{eq:r_V}, \eq{eq:A_V} and \eq{eq:lam_V} the energy density $\lambda$ becomes
\bea
	\frac{8\pi G}{c^4} \lambda (V,\alpha ) 
		&=& \fr{V^2}{\alpha ^2 (1+V)+V^2}\fr{V^2(V- v_2)^{\varepsilon -1}}{(V- v_1)^{\varepsilon +1}}\;.\label{eq:lam_V_lim}
\eea
where $v_1\simeq 1$, $v_2\simeq -\alpha ^2+1$, $\varepsilon  \simeq 1$ for a large $\alpha$, 
and \eq{eq:lam_V_lim} implies 
\bea
	\begin{array}{ll}
	\underset{\alpha \rightarrow \infty}{\lim }\frac{8\pi G}{c^4} \lambda = 0& \text{(the Schwarzschild solution)} \\
	\underset{V\rightarrow \infty}{\lim}\fr{8\pi G}{c^4} \lambda = 1 &\text{(different solution near}\;r=1)
	\end{array}
\eea
depending on the order of taking the limits. This implies that there is no smooth function connecting our solution to the Schwarzschild solution as we also see in the figure.
The figure also exhibits that there is a minimum area radius $r=1$.

\subsection{PPN parametrization}\label{sec:ppn}
 As we have seen in the previous subsection, the spherically symmetric solution\footnote{We have taken the $V>0$ branch.} predicted by our new theory of gravitation deviates from the Schwarzschild solution, the amount of which gets smaller for larger $(a+1)$.
 To quantify the difference from general relativity, it is standard to use the PPN parametrization~\cite{Will} in the isotropic coordinates,
 \bea
ds^2=-B(\rho)c^2dt^2+K(\rho)[d\rho^2+\rho^2d\theta^2+\rho^2\sin ^2 \theta d\phi ^2],
\eea 
where $B(\rho)$ is as defined before and $K(\rho)$ is a function of the radius $\rho$ given by
\bea
\sqrt{K}d\rho=\sqrt{A}dr\;,\\
\sqrt{K}\rho=r.
\eea 
We see that
\bea
K=r^2\exp[-2\int \sqrt{A}\frac{dr}{r}]\;,\\
\rho=\frac{r}{\sqrt{K}}=\exp[\int\sqrt{A}\frac{dr}{r}].
\eea 

Recalling the expression for $r$ in terms of the parameter $V$\footnote{Since $V>0$, the absolute value symbol can be removed.},
\bea
r
	&=&\frac{(V- v_1)^{(\varepsilon+1)/2}}{(V- v_2)^{(\varepsilon-1)/2}V},
\label{r/V} 
\eea 
we notice that the limit of spatial infinity $r\rightarrow \infty$ corresponds to $V\rightarrow 0$. We first express $K$ and $\rho$ as functions of $V$ and then Taylor expand them in powers of $V$ to see the asymptotic behavior of $K$ and $B$ at $r\rightarrow \infty$.
Take the logarithmic derivative of ($\ref{r/V}$) to obtain 
\bea
\frac{dr}{r}
	&=&\frac{dV}{V}[\frac{\varepsilon+1}{2}\frac{V}{V-v_1}-\frac{\varepsilon-1}{2}\frac{V}{V- v_2}-1]=-\frac{dV}{V}\frac{\alpha^2}{(V- v_1)(V- v_2)}
\eea 
and
\bea
Q:=\int \sqrt{A}\frac{dr}{r}
	&=& 
	-\int \frac{dV}{V}\frac{1}{(1-\frac{V}{ v_1})^{1/2}(1-\frac{V}{ v_2})^{1/2}}. 
\eea 
We expand $Q$ in powers of $V$ as
\bea
Q=-\log V+ V/2+\dots
\eea 
to find the Taylor expansions of  $K\rho^2=r^2=\frac{(V- v_1)^{\varepsilon+1}}{(V- v_2)^{\varepsilon-1}V^2}$ and $\rho=e^{Q}$.
The result is
\bea
\rho=V^{-1}(1+\frac{1}{2}V+\dots).
\eea 
Inverting this equation for $V$,we obtain
\bea
V=\frac{1}{\rho}+\frac{1}{2\rho^2}+\dots.
\eea
Inserting this into the Taylor expansions of $B$ and $r^2$, we arrive at the asymptotic expansions
\bea
B
&=&|\frac{ v_1}{ v_2}|^{\varepsilon}[1-\frac{1}{\rho}+\frac{1}{2\rho^2}+\dots]
\propto (1-2U+2\beta U^2+\dots),\\
K
 &=&\ \frac{|v_1|^{\varepsilon+1}}{|v_2|^{\varepsilon-1}}[1+\frac{1}{\rho}+\dots],
\propto 1+2\gamma U+\dots,
\eea
where $U=\frac{1}{2\rho}$ is the Newtonian potential.
Comparing with the standard PPN parametrization, we see from the coefficient of the quadratic term in the expression for  $B$ that the PPN parameter $\beta =1$ and the linear term in the expression for $K$ that the PPN parameter $\gamma=1$
and therefore the result coincides with the result of general relativity. The deviation appears only in the quadratic term in the expression for  $K$, which would give the lower limit of $\alpha$.

\section{Summary}\label{sec:summary}

Postulating the minimal time to send a bit of information in the Einstein synchronization of the two clocks located at different positions, we have introduced the  metric extended to the information space, where the important ingredient is the unit time-like velocity 
field. The extension of the metric changes the red shift formula while the geodesic equation is kept intact. Extending the gauge symmetry of the metric to the 5-dimensional general invariance, we start with the Einstein-Hilbert action in the 5-dimensional space. After the 4+1 decomposition of the 5-dimensional Einstein-Hilbert action we arrive at the effective action which includes the Einstein-Hilbert action for gravity, the Maxwell-like action for the velocity field and the Lagrange multiplier term which ensures the normalization of the time-like velocity field. As an application, we investigated a solution of the field equations in the case that a 4-dimensional part of the extended metric is spherically symmetric.We have found that the Birkhoff theorem holds. In the GR limit, the solution approaches the Schwarzschild space-time in the weak field regime, while  the space-time is significantly different from the Schwarzschild space-time near the minimum radius.

\section{Discussion}\label{sec:discussion}
\subsection{Remarks}
Let us look at the informational theory of relativity (ITR) from the view point of the Einstein-Aether (EAE) theory developed by Jacobson and co-workers\cite{EAE}. The effective Lagrangian of EAE for the metric $g_{\mu\nu}$ and the vector field $u^ {\mu}$ reads
\bea
	&&L=R-K^{\mu\nu}_{\alpha\beta}D_{\mu}u^{\alpha}D_{\nu}u^{\beta}+ \lambda(g_{\alpha\beta}u^{\alpha}u^{\beta}+1)\;,\\
	&&K^{\mu\nu}_{\alpha\beta}= c_{1}g^{\mu\nu}g_{\alpha\beta}+ c_{2}\delta^{\mu}_{\alpha } \delta^{\nu}_{\beta}+c_{3}\delta^{\mu}_{\beta} \delta^{\nu}_{\alpha}+c_{4}u^{\mu}u^{\nu}g_{\alpha\beta},
\label{EAE}
\eea
where ${c _\alpha}'s$ are unspecified parameters.
One may naturally suspect that the above Lagrangian reduces to
our effective Lagrangian \eq{action} if we impose further symmetry. Actually this is the case. 
Comparing (\ref{EAE}) and (\ref{Stotal01}), we see after a simple algebra that
\bea
&&2c_1=-c_2=2c_3=\frac{1}{a+1},\;\;\; c_4=0.
\eea
Using the result of \cite{EAE}, we can see some characteristic properties of our theory.  The spin-2, spin-1 and spin -0 wave velocities squared are given by $\frac{v_{\rm tensor}^2}{c^2}=\fr{1}{1-2c_1}=\frac{v^2_{\rm vector}}{c^2}$ and $v^2_{\rm scalar}=0$, repesectively. The wave velocity of the massless matter field $\psi$ coincides with $\frac{v^{2}_{matter}}{c^2}=\fr{1}{1-2c_1}$. This can be also explained by the particle picture. A massless particle in 5-dimensions satisfies the constraint: $\hat{g}^{\hat{\mu}\hat{\nu}}P_{\hat\mu} P_{\hat\nu}=0$
for the 5-momentum $P_{\hat\mu}$. Assuming that the particle does not go into the information space, we see that $P_4=0$ and therefore $\hat{g}^{\mu\nu}P_{\mu} P_{\nu}=0$ holds. For the flat space-time this reduces to $(-1+\frac{1}{a+1})(P_0)^2+{\bf P}^2=0$, i.e., the corresponding phase velocity $v$ of the massless field is given by $\frac{v^2}{c^2}:=\frac{ (P_0)^2}{{\bf P}^2}=\frac{1}{1-\frac{1}{a+1}}$ from the de Broglie-Enistein relation. Note that $2c_1:=\fr{1}{a+1}$ is very small. However small it would be, the deviation of the unique velocity of the spin 2, 1 and massless matter fields from the light velocity c is puzzling from the Einstein-Aether theoretical point of view. From the view point of ITR, it is simply an artifact of the observed time rather than the true time as discussed in Sect.2. Suppose the massless field $\psi$ is a function of the true time $t-\tau \sigma$. Then $P_0+\fr{P_4}{\tau}=0$ holds so that the true phase velocity is simply given by $\frac{{v_{true}}^2}{c^2}:=\frac{ (P_0)^2}{{\bf P}^2}=1$.

The gauge symmetry seemingly lacks intuition in the EAE theory but in our informational approach, the symmetry comes from the freedom to choose the origin of the information coordinate $\sigma$ at each space-time point.
  As for another alternative gravity theory TeVeS, it differs from ours, because it does not contain the term quadratic in the velocity field and linear in the Ricci tensor. The scalar field which exists in TeVeS will emerge also in our theory if the scale of the information space allows to depend on space-time coordinates $x^\mu $, that is $a(x)$.

We can immediately observe that the Minkowski space-time with
$u^{\mu}=(1,0,0,0)$ is a particular solution. This fact is an assuring
evidence because we always assume that the space-time is locally Minkowskian.

We have obtained the geometry of the outer space $r>1$ in some unit as a spherically symmetric solution. A natural question is what is likely
the geometry of the inner space. At the moment, we have the two
possibilities in our mind: (i)there is an analytic coordinate similar to the Kruskal coordinate in GR (ii) there may be a solution $u^{r}\neq 0$ in general. 
Note that the non-vanishing $ u^{r}$ cannot be gauged away by
the gauge transformation (\ref{u})(\ref{eq:gauge_g}) without affecting the metric ansaz (\ref{SPmetric}), specifically $g_{tr}=0$.
We have a plan to explore a global solution defined in the whole space-time and see whether the solution corresponds to a black hole.
\subsection{Further discussions: possible origin of $\sigma$}\label{sec:orgin_sigma}
Here we present a model of the synchronization of all the clocks in a domain of the universe for illustration to see how the randomness and the new coordinate $\sigma$ appear. Of course, this is only one of possible models and the contents of the main text does not depend on the details of the present model. 


Fix the number of bits $n$ to describe the coordinate difference $dx_{true}^\mu=x_{true}^\mu(2)-x_{true}^\mu(1)$ and the information of the light signal\footnote{The bits for the difference of the coordinates of the nearby two points
need less bits for each coordinate. For example,$t(1)=2018,30713160901,t(2)=2018,30713160913$
need 15 digits while their difference needs only 2 digits $12$.
}.
  In this paper we assume that the fixed number $n$ is sufficiently large so that we can treat the coordinates as continuous valued to a good approximation, following the standard argument
in physics and information science as well.
 Let the compression rate of the signal be $\sigma$ 
so that
the $n$ bits of signal is compressed to $[ \sigma n ]$ bits of signal and we write it as $\sigma n$ because the difference is negligible for a sufficiently large $n$.
 As stated before, the time needed to send the signal is $\tau_0\times\sigma n$,  where the $\tau_0$ is
a fundamental constant to send one bit of information. The time $\tau=\tau_0 n$, 
can be regarded as a time scale of the physics under consideration. Our theory depends on the time scale $\tau$ but not the fundamental constant $\tau_0$ explicitly as we will
subsequently see. 
 We need the exchange of the light signal for the clock synchronization, so that we begin with a process to send the signal in the minimal time.

For each synchronization of a pair of clocks,  a single $\sigma$ is assigned as a compression rate of the information of the starting time $t_0$ of the synchronization process. To synchronize all the clocks in a domain of the universe we need many exchanges of the light signals and therefore many $\sigma$'s, each of which depends on arbitrarily chosen starting time of the synchronization process.Therefore, the value of $\sigma$ is stochastic, when we consider all the observed events in the universe. 
   
In a field theory every field is a function of observed 
coordinates corresponding to an event as indicated by a dot p for 
example in the \fig{fig:a1}. To define a local event in our context we 
conveniently decompose \fig{fig:a1} into \fig{fig:a2} with the understanding that 
the two events of the same values of $\sigma$ in \fig{fig:a2} is linked in 
\fig{fig:a1}.
\begin{figure}[h]
\begin{center}
\includegraphics[width=4cm,clip]{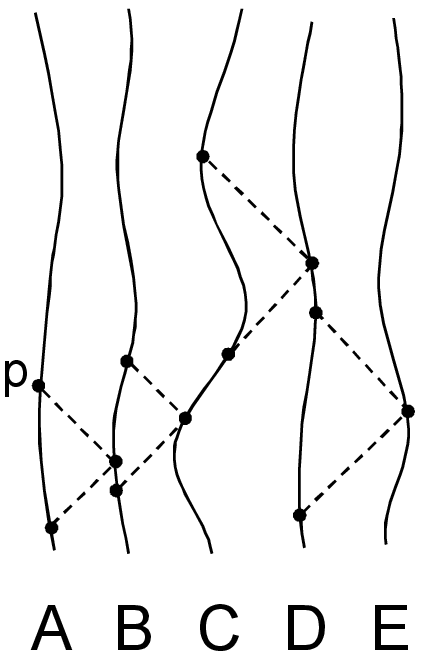}
\end{center}
\caption[communication and events]{}
\label{fig:a1}
\end{figure} 
\begin{figure}[h]
\begin{center}
\includegraphics[width=4cm,clip]{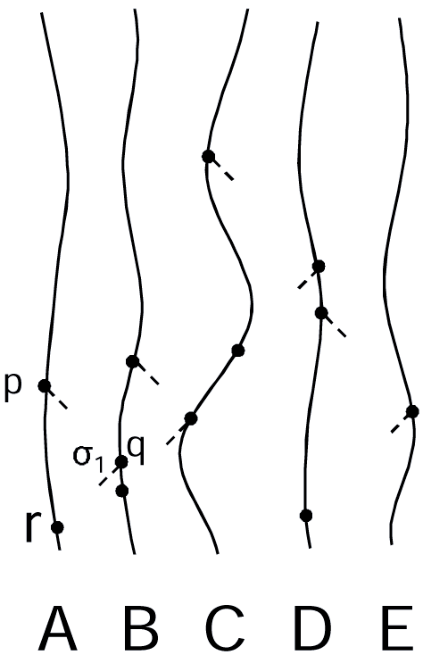}
\end{center}
\caption[decomposed events]{ }
\label{fig:a2}
\end{figure} 
 Now we are going to explain how the randomness appears in the signal exchange and then introduce the variable $\sigma$ as the compression rate of the signal information.

Consider the two nearby events (1) and (2) which are not necessarily connected by the light signal. (See \fig{fig:a3}). Here we can introduce the new information space $\sigma $ as a new variable independent of the space-time coordinates.
Now $\sigma(1)$ and  $\sigma(2)$ can be different in general. The coordinate difference $dx_{true}^\mu=x_{true}^\mu(2)-x_{true}^\mu(1)$ is therefore rewritten as  $dx_{true}^\mu=x_{obs}^\mu(2)-x_{obs}^\mu(1)-(\sigma(2)-\sigma(1))\tau u^{\mu}$. In short, 
 $dx_{true}^\mu=dx_{obs}^\mu-d\sigma\tau u^{\mu}$. 
 Note that $u^{\mu}$ is common for the spatial-temporal  domain near the point  p , because  the value of  $u^{\mu}$ is the same at the three points  p, q and r for the Einstein synchronization. The difference of the  $u^{\mu}$ is small of  the second order of the distance between the two world lines in the neighborhood of p,q and r in \fig{fig:a3}.
 
According to the Shannon optimal compression theorem\cite{Shannon}, the optimal compression rate is generally given by $\sigma_0=-\frac{\log P(t_0)}{n}$ for an event probability $P(t_0)$ where $t_0$ is the initial light transmitting time. In our particular  case  the event is the start of  the synchronization and its time $t_0$ coded by $n$ bits is compressed to $\sigma_0 n$. The metric in the $\sigma$ space is simply given by $(d\sigma)^2$. On average, it becomes the well-known Fisher metric $P(d\sigma)^2=\frac{(dP)^2}{n^{2}P}$. We adopt the metric $(d\sigma)^2$ for
 our particular stochastic synchronization process. Recall that the stochaticity comes from the random choice of $t_0$.

On the other hand the causality is primarily defined in the true coordinates $x_{true}^\mu$, which can be inferred through the observed coordinates $x_{obs}^\mu$ given the stochastic variable $\sigma$  through the relation $x_{true}^\mu=x_{obs}^\mu-\tau \sigma u^\mu$.Therefore, we have to treat the $\sigma$ as a variable independent of the observed coordinates $x_{obs}^\mu$. Putting this more illustrative, we may imagine the picture like
\fig{fig:a3}.

\begin{figure}[h]
\begin{center}
\includegraphics[width=4cm,clip]{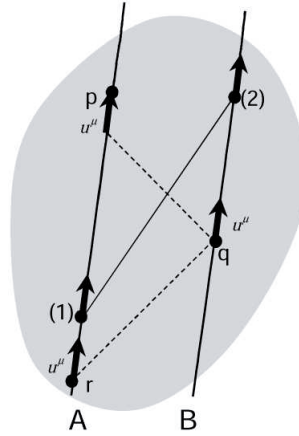}
\end{center}
\caption[]{The observer A confirms the synchronization of the two clocks at A and B by exchanging the
light signals. At p and q the value of the velocity field $u^{\mu}$ is the same and so is at q and r.
Therefore the field $u^{\mu}$ is constant in the neighborhood of the synchronization region. }
\label{fig:a3}
\end{figure} 
  
  One might wonder why the above coordinate transformation makes any physical difference from
 the standard general relativity. The point is that the field like the metric tensor $g_{\mu\nu}$
 is measured at the point of the observed coordinates $x_{obs}^\mu$. 
 Therefore, the metric tensor is primarily defined as a function of the observed coordinate 
 $x_{obs}^\mu$. Explicitly, the metric is given by
 $$
 ds_{grav}^2=g_{\mu\nu}(x_{obs})dx_{true}^\mu dx_{true}^\nu
 $$
 for the space-time part. At this stage, we have already shifted the geometrical picture from the operational picture as we historically did in the introduction of the metric in the relativity.

We take the information part of the metric as,
 $$
 ds_{inf}^2=(a+1)(d\sigma)^2
 $$
 with a positive parameter $(a+1)$, which is to be determined by experiments.
 For the whole metric,it is natural to add the two of them a l{\'a} Pythagoras as described in the main text. By averaging the total metric $ds_{total}^2=ds_{grav}^2+ds_{inf}^2$ with the probability $P$,
 the informational part would become the Fisher metric. 
 Our discussion on the origin of $\sigma $
is the stochasticity of the time $t_1$ which comes from the arbitrariness of the time $t_0$ in the limited space-time domain.

\subsection{Outlook}
 The coordinate $\sigma$ of the information space is
introduced as bits of the arrival time information of Bob and
$\tau d\sigma$ is the minimum time to send the amount of
$d\sigma$ information. One may think of the Shannon compression to minimize the message and therefore the time to
send it fixing the accuracy of time information. At the moment, we do not know any algorithm of the compression for a given space-time point of Bob. We suspect that
it would not be deterministic but rather stochastic because the value of the compression factor i.e., the Shannon entropy would sensitively depend on the exact location of Bob relative to the nearby clock. Note that the Shannon compression gives a fractional bit of information rather than integer. We admit that the physical characteriztion of the informational
coordinate $\sigma$  is yet to be clarified. However, the result of the present work remains valid since we use it only
operationally.

 Obviously the Friedmann-Robertson-Walker like cosmological model and the Kerr like rotating solutions are most interesting. It remains to be seen whether ITR can explain the astronomical effects which have been normally attributed to the dark energy and dark matter, while TeVeS is motivated by the modification of gravity theory without introduction of dark matter.

There remain many open questions which are at the moment far reaching for the present authors.
\begin{enumerate}
\item Can we experimentally measure the value of $\tau?$
Is it related to quantum mechanics?
\item How can we detect the vector wave in principle?
\item Even far reaching, how can we go over to quantum gravity?
\end{enumerate}
\appendix

\section{Outline of Einstein's derivation of Lorentz transformation}\label{app:A}
For the light propagation parallel to the direction of the moving frame velocity, we see $ \Delta t_1=\fr{\ell}{c-v}$ for the time interval of the light propagation from the clock {\bf a} to the clock {\bf b} which both co-move with the moving frame keeping the distance $\ell$ between them, and $\Delta t_2=\fr{\ell}{c+v}$ for that of the  light propagation in the opposite direction from the clock {\bf b} to the clock {\bf a}. 

Introducing $\widetilde{x}:=x-vt$ which is the rest frame relative distance viewed from the position $vt$ of the origin of the moving frame and the moving frame time $\eta$ is in general a function of the rest frame coordinates $(t,\widetilde{x},y,z)$. The Einstein synchronization condition in the moving frame becomes 
\bea
	\fr{1}{2} \left ( \eta(t,0,0,0)+\eta(t+\Delta t_1+\Delta t_2, 0,0,0) \right )=\eta (t+\Delta t_1,\ell,0,0) \;.\label{app:deta_x}
\eea
In the linear approximation, we obtain
\bea
	\fr{\pa \eta}{\pa t}+\fr{c^2-v^2}{c^2}\fr{\pa \eta}{\pa \widetilde{x}}=0 \;.
\eea

On the other hand, for the light propagation perpendicular to the direction of the moving frame velocity, we see $\Delta t_3=\fr{\ell}{\sqrt{c^2-v^2}}$ for the time interval of the light propagation from the clock {\bf a} to the clock {\bf b} (or from the clock {\bf b} to the clock {\bf a}), sitting in the relative position perpendicular to the direction of the moving frame velocity. In general from the origin to the position $y$ of the $y$ coordinate, the time interval $t$ measured with the rest frame time can be written as
\bea
	&&t=\fr{y}{\sqrt{c^2-v^2}}\;. \label{app:LP_perp}
\eea
From the synchronization condition in the moving frame, $\fr{1}{2} \left ( \eta(t,0,0,0)+\eta(t+2\Delta t_3,0,0,0) \right )=\eta (t+\Delta t_3,0,\ell,0) $ in the linear approximation, we obtain $\fr{\pa \eta}{\pa y}=0\label{app:deta_y}$
and the similar relation for the $z$ coordinate. 
 
Now let us briefly see how to derive the Lorentz transformation.
Assume the linearity of $\eta$ in the rest frame coordinates
\bea
	\eta =A(v) (t-\fr{\widetilde{x}}{c^2-v^2})+E\;, \label{app:Aeta}
\eea 
where $A$ is in general a function of $v$ and $E$ is a constant. For simplicity we set the space origin of the rest frame and moving frame are both in the same position initially, that is $x=0$ and $X=0$ at the time $t=0$ and $\eta =0$, and therefore we set $E=0$ in \eq{app:Aeta} in the following discussion. Imposing the invariance of light velocity, the light trajectory observed in the moving frame can be expressed as
\bea
	X=Y=Z=c\eta\;. \label{app:LP_inv}
\eea   
From using $t =\fr{\widetilde{x}}{c-v}  \label{app:LP_para}$ and Eqs.~(\ref{app:Aeta}) and (\ref{app:LP_inv}), for the light parallel to the moving frame motion we have
\bea
	X=A(v)\fr{c^2}{c^2-v^2}\widetilde{x}\;. \label{app:Xx}
\eea

From Eqs.(\ref{app:LP_perp}), (\ref{app:Aeta}) and (\ref{app:LP_inv}), for the light perpendicular to the moving frame motion we have
\bea
	Y=A(v)\fr{cy}{\sqrt{c^2-v^2}}\; \label{app:Yy}
\eea
and the similar relation for the $Z$ coordinate. Clearly the Lorentz transformation by $v$ after the Lorentz transformation by $-v$ is the identity transformation. We obtain
\bea
	&&A(v)A(-v)\fr{1}{1-\fr{v^2}{c^2}}=1\;.\label{app:AA}
\eea
Note that for the light propagation perpendicular to the moving frame velocity, the relative distance between the clock {\bf a} and the clock {\bf b} viewed from the rest frame is independent of the direction of the moving frame velocity, so that $A(v)=A(-v)$
, which implies 
\bea
		&&A(v)=\sqrt{1-\left ( \fr{v}{c}\right )^2}\;, \label{app:Avc}
\eea
Then from Eqs.~(\ref{app:Aeta}),(\ref{app:Xx}),(\ref{app:Yy}) and (\ref{app:Avc}) we finally find the Lorentz transformation     
\bea
	&&\eta =\fr{t-\fr{v}{c^2}x}{\sqrt{1-\left ( \fr{v}{c} \right )^2}} 
	\;,X=   \fr{x-vt}{\sqrt{1-\left (\fr{v}{c} \right )^2}} \;, \\
	&&Y= y \;, Z= z\;.
\eea

\section{Christoffel symbols, Ricci, Ricci scalar, Einstein tensor}

 The computation in section {\ref{sec:classic} and subsection {\ref{sec:action}}} is rather straightforward but seemingly unaccustomed to many readers so that we decided to explicitly  write out the details.
$\hat{\nabla} _{\hat{\mu}}$ denotes the $5$-dimensional covariant derivative with respect to the information-space-time metric $\hat{g}_{\hat{\mu}\hat{\nu}}$ and
$D_\mu$ denotes the 4-dimensional covariant derivative with respect to the $4$-dimensional space-time metric $g_{\mu \nu}$.\\
With a scale of information space, $a \tau^2$, the metric and the inverse metric are respectively given by 
\bea
	\hat{g}_{\hat{\mu}\hat{\nu}}
		&=& \left (
			\begin{array}{cc}
				g_{\mu\nu} 	& -\tau u_\mu \\
				-\tau u_\nu &a \tau ^2
		    \end{array}
		    \right )\;, \label{met2} \\
	\hat{g}^{\hat{\mu}\hat{\nu}}
		&=& \left (
			\begin{array}{cc}
				g^{\mu\nu}+\fr{1}{a+1}u^\mu u^\nu 	& \fr{1}{a+1}\fr{1}{\tau}u^\mu \\
				\fr{1}{a+1}\fr{1}{\tau}u^\nu 		& \fr{1}{a+1}\fr{1}{\tau ^2}
		    \end{array}
		    \right )\;. \label{imet2}
\eea
From \eq{met2} and \eq{imet2}, the Christoffel symbol is calculated to be
 \bea
	&&\hat{\Gamma} ^\lambda _{\mu\nu}=\Gamma ^\lambda _{\mu\nu} -\fr{1}{2}\fr{1}{a+1}u^\lambda S_{\mu\nu}\;,	\label{Chr-lmn}\\
	&&\text{where } S_{\mu\nu}:=D_\mu u_\nu +D_\nu u_\mu \\ 
	&&\hat{\Gamma} ^4_{\mu\nu}=-\fr{1}{2}\fr{1}{a+1}\fr{1}{\tau}S_{\mu\nu}\;,\label{Chr-4mn}\\
	&&\hat{\Gamma}^\lambda _{4\mu}=-\fr{1}{2}\tau (g^{\lambda \alpha}+\fr{1}{a+1}u^\lambda u^\alpha )A_{\mu \alpha}\;,\label{Chr-l4m} \\
	&&\text{where } A_{\mu \nu}:=D_\mu u_\nu -D_\nu u_\mu \\
	&&\hat{\Gamma}^\lambda _{4 \lambda }=0\;, \label{Chr-ll4}\\
	&&\hat{\Gamma}^\lambda _{44}=0 \;, \label{Chr-l44} \\
	&&\hat{\Gamma}^4_{\mu 4}=-\fr{1}{2}\fr{1}{a+1}u^\alpha A_{\mu\alpha} \;, \label{Chr-4m4} \\
	&&\hat{\Gamma}^\lambda _{44}=0 \label{l44}\;,\\
	&&\hat{\Gamma}^4_{44}=0\;. \label{Chr-444}
 \eea
 From \eq{Chr-l4m} and \eq{Chr-444}, we see that
 \bea
	&&\hat{\Gamma}^{\hat{\lambda}}_{4\hat{\lambda}}=0 \;. \label{Chr-H44}
 \eea
Let us denote the 5-dimensional covariant derivative as $\hat{\nabla} _{\hat{\mu}}$ satisfying $\hat{\nabla}_{\hat{\mu}}\hat{g}_{\hat{\alpha}\hat{\beta}}=0$, while
the 4-dimensional space-time covariant derivative as $D_\mu$, with $D_\mu g_{\alpha \beta}=0$.\\

In this paper we adopt the convention of the Ricci tensor as
\bea
 \hat{R}^{\hat{\alpha}}_{~\mu\hat{\alpha}\nu}
	=\pa _{\hat{\alpha}}\hat{\Gamma} ^{\hat{\alpha}}_{\mu\nu} 
	-\pa_\nu \hat{\Gamma} ^{\hat{\alpha}}_{\mu \hat{\alpha}}
	+\hat{\Gamma}^{\hat{\alpha}}_{\hat{\alpha}\hat{\beta}} 
     \hat{\Gamma}^{\hat{\beta}}_{\nu\mu}
    -\hat{\Gamma}^{\hat{\alpha}}_{\nu \hat{\beta}} 
     \hat{\Gamma}^{\hat{\beta}}_{\hat{\alpha}\mu} .
\eea
Then each component of the Ricci tensor is explicitly given by 
\bea
	\hat{R}_{44}=\hat{R}^{\hat{\lambda}}~_{4\hat{\lambda}4}
	&=&-\uf \Gamma ^{\hat{\alpha}} _{4\hat{\beta}} \uf \Gamma ^{\hat{\beta}}_{\hat{\alpha}4} \nonumber \\
	&=&-\fr{1}{4}\tau ^2 \hat{g}^{\alpha \lambda }\hat{g}^{\beta \rho}A_{\alpha \rho}A_{\beta \lambda} \nonumber\\
    &=&\fr{1}{4}\tau^2 (A_{\alpha \beta}A^{\alpha \beta}+\fr{2}{a+1}u^\alpha u^\beta A_{\rho \alpha}A^\rho ~_\beta )\;,\label{tensor:R44}\\
	\hat{R}_{4\mu}=\hat{R}^{\hat{\lambda}}~_{4\hat{\lambda}\mu}
	&=&-\fr{1}{2}\tau D_\lambda (\hat{g}^{\lambda \rho}A_{\mu\rho})
		+\fr{1}{2}\tau\fr{1}{a+1}u^\alpha \bigl ( (D_\beta u_\alpha)D_\mu u^\beta -(D_\alpha u_\beta )D^\beta u_\mu \bigr ) \nonumber\\
		&&+\fr{1}{2}\tau\fr{1}{(a+1)^2}u^\alpha u^\beta u^\lambda  \bigl ( (D_\alpha u_\beta )A_{\mu\lambda } \bigr ) \;,\label{tensor:R4m} \\
	\hat{R}_{\mu\nu}=\hat{R}^{\hat{\lambda}}~_{\mu \hat{\lambda}\nu}
	&=&R_{\mu\nu}
		-\fr{1}{2}\fr{1}{a+1}D_\alpha \bigl (u^\alpha (D_\mu u_\nu+D_\nu u_\mu )+\fr{1}{a+1}D_\nu \bigl (u^\alpha D_\mu u_\alpha \bigr ) \nonumber\\
	&&-\fr{1}{(a+1)^2}u^\alpha u^\beta (D_\mu u_\alpha )D_\nu u_\beta 	
		-\fr{1}{2}\fr{1}{a+1}\bigl ((D_\nu u^\lambda )D_\mu u_\lambda -(D_\lambda u_\nu)D_\lambda u_\mu ) \bigr )\;,\label{tensor:Rmn}
\eea 
and the Ricci scalar is
\bea
\hat{R}&=&\hat{g}^{\hat{\alpha}\hat{\beta}}\hat{R}_{\hat{\alpha}\hat{\beta}}
		=\hat{g}^{44}\hat{R}_{44} +\hat{g}^{\alpha 4}\hat{R}_{\alpha 4}+\hat{g}^{4\beta}\hat{R}_{4\beta}
			+\hat{g}^{\alpha \beta}\hat{R}_{\alpha\beta} \;,  \nonumber\\
		&=&\fr{1}{a+1}\fr{1}{\tau ^2} \hat{R}_{44}+2\fr{u^\beta}{(a+1)\tau} \hat{R}_{4\beta}+(g^{\alpha\beta}+\fr{1}{a+1}u^\alpha u^\beta ) \hat{R}_{\alpha\beta} \;.\label{RS-01}
\eea
Substituting \eq{tensor:R44}, \eq{tensor:R4m}, \eq{tensor:Rmn} into \eq{RS-01} we have
\bea
	\hat{R}
		&=&R+\fr{1}{a+1}R_{\alpha \beta}u^\alpha u^\beta \nonumber \\
		&&+\fr{1}{4}\fr{1}{a+1}(g^{\alpha \beta}+\fr{2}{a+1}u^\alpha u^\beta )A_{\alpha \rho}A_{\beta}^{~~\rho}
		-\fr{1}{a+1}u^\alpha D_\lambda \bigl [(g^{\rho \lambda}+\fr{1}{a+1}u^\rho u^\lambda )A_{\alpha \rho} \bigr ] \nonumber\\
		&&+\fr{1}{a+1}(g^{\alpha \beta}+\fr{1}{a+1}u^\alpha u^\beta ) 
			\Bigl [-D_\rho (u^\rho D_\alpha u_\beta )+D_\beta (u^\rho D_\alpha u_\rho )-\fr{u^\rho u^\sigma}{a+1} (D_\alpha u_\rho )D_\beta u_\sigma \nonumber\\
		&&			+\fr{1}{2} \bigl \{ (D_\lambda u_\alpha )D^\lambda u_\beta -(D_\alpha u_\lambda )D_\beta u^\lambda \bigr \} \Bigr ]\;, \label{App:RicSc} 
\eea
or alternatively,
\bea
	\hat{R}
		&=&\hat{g}^{\alpha\beta}R_{\alpha\beta} \nonumber \\
		&&+\fr{1}{4}\fr{1}{a+1}\hat{g}^{\alpha \rho}\hat{g}^{\beta \sigma}A_{\alpha \rho}A_{\beta\sigma}
		-\fr{1}{a+1}u^\alpha D_\lambda (\hat{g}^{\lambda\rho}A_{\alpha \rho} ) \nonumber \\
		&&+\fr{1}{a+1}\hat{g}^{\alpha \beta} 
			\Bigl [-D_\rho (u^\rho D_\alpha u_\beta )+D_\beta (u^\rho D_\alpha u_\rho )-\fr{u^\rho u^\sigma}{a+1} (D_\alpha u_\rho )D_\beta u_\sigma \nonumber\\
		&&			+\fr{1}{2} \bigl \{ (D_\lambda u_\alpha )D^\lambda u_\beta -(D_\alpha u_\lambda )D_\beta u^\lambda \bigr \} \Bigr ]\;. \label{App:RicSc2} 
\eea
\section{Metric Determinant}\label{app:determinant}
   Using the determinant expansion method, the 5-dimensional metric $\det (\hat{g}_{\hat{\mu}\hat{\nu}})$ can be expanded directly.
\bea
	\det (\hat{g}_{\hat{\mu}\hat{\nu}})
		&=&\det \left ( 
				\begin{array}{c|c}
				g_{\mu\nu} 	&-\tau u _\nu \\
				\hline
				-\tau u_\mu 		&a\tau ^2
				\end{array}
				\right ) 
	=\det \left ( 
				\begin{array}{ccccc}
				g_{00} 	&g_{01}	&g_{02}	&g_{03}	&-\tau u _0 \\				
				g_{10} 	&g_{11}	&g_{12}	&g_{13}	&-\tau u _1 \\
				g_{20} 	&g_{21}	&g_{22}	&g_{23}	&-\tau u _2 \\
				g_{30} 	&g_{31}	&g_{32}	&g_{33}	&-\tau u _3 \\
				-\tau u_0&-\tau u_1 &-\tau u_2 &-\tau u_3 &a\tau ^2
				\end{array}
				\right ) \nonumber \\
				&&
			 \nonumber	\\
				&=&
					(-1)^{0+4}(-\tau u_0)\det
					\left ( 
					\begin{array}{cccc}
					g_{01} 	&g_{02}	&g_{03}	&-\tau u _0 \\				
					g_{11} 	&g_{12}	&g_{13}	&-\tau u _1 \\
					g_{21} 	&g_{22}	&g_{23}	&-\tau u _2 \\
					g_{31} 	&g_{32}	&g_{33}	&-\tau u _3 \\
					\end{array}
					\right )
					+(-1)^{1+4}(-\tau u_1)\det
					\left ( 
					\begin{array}{cccc}
					g_{00} 	&g_{02}	&g_{03}	&-\tau u _0 \\				
					g_{10} 	&g_{12}	&g_{13}	&-\tau u _1 \\
					g_{20} 	&g_{22}	&g_{23}	&-\tau u _2 \\
					g_{30} 	&g_{32}	&g_{33}	&-\tau u _3 \\
					\end{array}
					\right ) \nonumber \\
				&&+(-1)^{2+4}(-\tau u_2)\det
					\left ( 
					\begin{array}{cccc}
					g_{00} 	&g_{01}	&g_{03}	&-\tau u _0 \\				
					g_{10} 	&g_{11}	&g_{13}	&-\tau u _1 \\
					g_{20} 	&g_{21}	&g_{23}	&-\tau u _2 \\
					g_{30} 	&g_{31}	&g_{33}	&-\tau u _3 \\
					\end{array}
					\right ) 
					+(-1)^{3+4}(-\tau u_3)\det
					\left ( 
					\begin{array}{cccc}
					g_{00} 	&g_{01}	&g_{02}	&-\tau u _0 \\				
					g_{10} 	&g_{11}	&g_{12}	&-\tau u _1 \\
					g_{20} 	&g_{21}	&g_{22}	&-\tau u _2 \\
					g_{30} 	&g_{31}	&g_{32}	&-\tau u _3 \\
					\end{array}
					\right )\nonumber \\
		&&					
					+a\tau ^2 \det
					\left ( g_{\mu\nu}
					\right )		
					\label{det_g_0}.
\eea 
We see the first term:
\bea
\nonumber
	(-1)^{0+4}(-\tau u_0)\det
					\left ( 
					\begin{array}{cccc}
					g_{01} 	&g_{02}	&g_{03}	&-\tau u _0 \\				
					g_{11} 	&g_{12}	&g_{13}	&-\tau u _1 \\
					g_{21} 	&g_{22}	&g_{23}	&-\tau u _2 \\
					g_{31} 	&g_{32}	&g_{33}	&-\tau u _3 \\
					\end{array}
					\right ) 
	&=&(-1)^{0+4}(-\tau u_0)\det
					\left ( 
					\begin{array}{cccc}
					g_{00} 	&g_{01}	&g_{02}	&g_{03} \\				
					g_{10} 	&g_{11}	&g_{12}	&g_{13} \\
					g_{20} 	&g_{21}	&g_{22}	&g_{23} \\
					g_{30} 	&g_{31}	&g_{32}	&g_{33} \\
					\end{array}\right )
					\det
					\left ( 
					\begin{array}{cccc}
					0 	&0	&0	&-\tau u ^0 \\				
					1 	&0	&0	&-\tau u ^1 \\
					0 	&1	&0	&-\tau u ^2 \\
					0 	&0	&1	&-\tau u ^3 \\
					\end{array}
					\right ) \nonumber \\
					\nonumber\\
	&=&(-1)^{0+4}(-\tau u_0)		
					\det \left ( g_{\mu\nu}	\right )
					(-1)^{1+0}(-\tau u^0) \nonumber \\					
	&=&-\tau^2 u_0u^0	\det \left ( g_{\mu\nu}	\right ),			
	\label{det_g_1}
\eea
the second term:
\bea
	(-1)^{1+4}(-\tau u_1)\det
					\left ( 
					\begin{array}{cccc}
					g_{00} 	&g_{02}	&g_{03}	&-\tau u _0 \\				
					g_{10} 	&g_{12}	&g_{13}	&-\tau u _1 \\
					g_{20} 	&g_{22}	&g_{23}	&-\tau u _2 \\
					g_{30} 	&g_{32}	&g_{33}	&-\tau u _3 \\
					\end{array}
					\right ) 
		&=&(-1)^{1+4}(-\tau u_1)\det
					\left ( 
					\begin{array}{cccc}
					g_{00} 	&g_{01}	&g_{02}	&g_{03} \\				
					g_{10} 	&g_{11}	&g_{12}	&g_{13} \\
					g_{20} 	&g_{21}	&g_{22}	&g_{23} \\
					g_{30} 	&g_{31}	&g_{32}	&g_{33} \\
					\end{array}
					\right )
					\det
					\left ( 
					\begin{array}{cccc}
					1 	&0	&0	&-\tau u ^0 \\				
					0 	&0	&0	&-\tau u ^1 \\
					0 	&1	&0	&-\tau u ^2 \\
					0 	&0	&1	&-\tau u ^3 \\
					\end{array}
					\right ) \nonumber \\
					\nonumber\\
	&=&(-1)^{1+4}(-\tau u_1)		
					\det \left ( g_{\mu\nu}	\right )
					(-1)^{0+0}(-\tau u^1) \nonumber \\					
	&=&-\tau^2 u_1u^1	\det \left ( g_{\mu\nu}	\right ),			
	\label{det_g_2}
\eea
the third term:
\bea
	(-1)^{2+4}(-\tau u_2)\det
					\left ( 
					\begin{array}{cccc}
					g_{00} 	&g_{01}	&g_{03}	&-\tau u _0 \\				
					g_{10} 	&g_{11}	&g_{13}	&-\tau u _1 \\
					g_{20} 	&g_{21}	&g_{23}	&-\tau u _2 \\
					g_{30} 	&g_{31}	&g_{33}	&-\tau u _3 \\
					\end{array}
					\right ) 
				&=&(-1)^{2+4}(-\tau u_2)\det
					\left ( 
					\begin{array}{cccc}
					g_{00} 	&g_{01}	&g_{02}	&g_{03} \\				
					g_{10} 	&g_{11}	&g_{12}	&g_{13} \\
					g_{20} 	&g_{21}	&g_{22}	&g_{23} \\
					g_{30} 	&g_{31}	&g_{32}	&g_{33} \\
					\end{array}
					\right )
					\det
					\left ( 
					\begin{array}{cccc}
					0 	&0	&0	&-\tau u ^0 \\				
					1 	&0	&0	&-\tau u ^1 \\
					0 	&1	&0	&-\tau u ^2 \\
					0 	&0	&1	&-\tau u ^3 \\
					\end{array}
					\right ) \nonumber \\
					\nonumber \\
				&=&(-1)^{2+4}(-\tau u_2)		
					\det \left ( g_{\mu\nu}	\right )
					(-1)^{0+0}(-1)(-\tau u^2) \nonumber \\
				&=&-\tau^2 u_2u^2	\det \left ( g_{\mu\nu}	\right ),		
	\label{det_g_3}
\eea
 and the fourth term:
\bea
	+(-1)^{3+4}(-\tau u_3)\det
					\left ( 
					\begin{array}{cccc}
					g_{00} 	&g_{01}	&g_{02}	&-\tau u _0 \\				
					g_{10} 	&g_{11}	&g_{12}	&-\tau u _1 \\
					g_{20} 	&g_{21}	&g_{22}	&-\tau u _2 \\
					g_{30} 	&g_{31}	&g_{32}	&-\tau u _3 \\
					\end{array}
					\right )
				&=&(-1)^{3+4}(-\tau u_3)\det
					\left ( 
					\begin{array}{cccc}
					g_{00} 	&g_{01}	&g_{02}	&g_{03} \\				
					g_{10} 	&g_{11}	&g_{12}	&g_{13} \\
					g_{20} 	&g_{21}	&g_{22}	&g_{23} \\
					g_{30} 	&g_{31}	&g_{32}	&g_{33} \\
					\end{array}
					\right )
					\det
					\left ( 
					\begin{array}{cccc}
					1 	&0	&0	&-\tau u ^0 \\				
					0 	&1	&0	&-\tau u ^1 \\
					0 	&0	&1	&-\tau u ^2 \\
					0 	&0	&0	&-\tau u ^3 \\
					\end{array}
					\right ) \nonumber \\
					\nonumber \\
				&=&(-1)^{3+4}(-\tau u_3)		
					\det \left ( g_{\mu\nu}	\right )
					(-1)^{0+0}(-\tau u^3) \nonumber\\					
	&=&-\tau^2 u_3u^3	\det \left ( g_{\mu\nu}	\right ).			
	\label{det_g_4}
\eea
Note that the relation $u_\mu=g_{\mu\nu}u^\nu$ is used to express cofactor matrices in the two simpler matrix products.
Substituting \eq{det_g_1}, \eq{det_g_2}, \eq{det_g_3} and \eq{det_g_4} into \eq{det_g_0}, we obtain
\bea
	\det (\hat{g}_{\hat{\mu}\hat{\nu}})=\tau ^2 (a -u_\xi u^\xi ) \det (g_{\mu\nu})\;.
\eea
\section{Simplifying action}\label{app:simplification}
   Let us simplify \eq{Stotal01}.
\bea
	S	&=&\fr{c^4}{16\pi G}\int \sqrt{u^\xi u_\xi \det (g_{\eta \iota})}d^4x\tau d\sigma \Bigl [\hat{g}^{\alpha\beta}R_{\alpha\beta} \nonumber \\
		&&+\fr{1}{4}\fr{1}{a+1}\hat{g}^{\alpha \rho}\hat{g}^{\beta \sigma}A_{\alpha \rho}A_{\beta\sigma}
		-\fr{1}{a+1}u^\alpha D_\lambda (\hat{g}^{\lambda\rho}A_{\alpha \rho} ) \nonumber \\
		&&+\fr{1}{a+1}\hat{g}^{\alpha \beta} 
			\Bigl \{ -D_\rho (u^\rho D_\alpha u_\beta )+D_\beta (u^\rho D_\alpha u_\rho )-\fr{u^\rho u^\sigma}{a+1} (D_\alpha u_\rho )D_\beta u_\sigma \nonumber \\
		&&			+\fr{1}{2} \bigl \{ (D_\lambda u_\alpha )D^\lambda u_\beta -(D_\alpha u_\lambda )D_\beta u^\lambda \bigr \} \Bigr \} +\frac{16 \pi G}{c^4}\lambda (g_{\mu\nu}u^\mu u^\nu +1) \Bigr ]\;.
		\label{a_Stotal01}
\eea 
    The third term in \eq{a_Stotal01} becomes,
\bea
-\fr{1}{a+1}u^\alpha D_\lambda (\hat{g}^{\lambda \rho}A_{\alpha \rho})
		&=&-\fr{1}{a+1}\fr{1}{\sqrt{u^\xi u_\xi \det g_{\eta \iota }}}
		\Bigl [D_\lambda \bigl ( u^\alpha \hat{g}^{\lambda \rho}A_{\alpha \rho}\sqrt{u^\xi u_\xi \det g_{\eta \iota }} \bigr ) \nonumber \\
		&&-(D_\lambda u^\alpha )\hat{g}^{\lambda \rho}A_{\alpha \rho}\sqrt{u^\xi u_\xi \det g_{\eta \iota }}
		  -u^\alpha \hat{g}^{\lambda \rho}A_{\alpha \rho}\fr{1}{2}(u^\xi u_\xi)^{-\fr{1}{2}}D_\lambda (u^\tau u_\tau ) \sqrt{\det g_{\eta \iota }}\Bigr ]\;. \nonumber \\
		&=&-\fr{1}{a+1}\Bigl [ -(D_\lambda u_\beta )g^{\beta \alpha}\hat{g}^{\lambda \rho}A_{\alpha \rho}
			-\fr{1}{2}\fr{u^\alpha \hat{g}^{\lambda \rho}A_{\alpha \rho}}{u_\xi u^\xi} D_\lambda (u_\tau u^\tau+1) \Bigl ] \label{t3_eq3}
\eea
The second term in \eq{t3_eq3} comes from the differentiation of $\sqrt{-\det \hat{g}_{\eta \iota}}$. Note that this term is proportional to $D_\lambda (u_\tau u^\tau +1)$, which can be partially integrated to  the same form as the constraint term in \eq{Sone} as
\bea
 &&\sqrt{-\det ( \hat{g}_{\eta \iota})}\fr{u^\alpha \hat{g}^{\lambda \rho}A_{\alpha \rho}}{u_\xi u^\xi} D_\lambda (u_\tau u^\tau +1)\nonumber \\
	&&=D_\lambda \Bigl [ \fr{u^\alpha \hat{g}^{\lambda \rho}A_{\alpha \rho} \sqrt{u_\delta u^\delta \det (g_{\epsilon \phi })}}{u_\xi u^\xi} (u_\tau u^\tau +1) \Bigl ] \nonumber \\	
	&&-\Bigl [ 
		\fr{1}{2} \fr{u^\alpha \hat{g}^{\lambda \rho}A_{\alpha \rho}}{u_\xi u^\xi} \sqrt{\fr{\det (g_{\epsilon \phi })}{u_\delta u^\delta }}D_\lambda (u_\delta u^\delta )
		+D_\lambda \bigl ( \fr{u^\alpha \hat{g}^{\lambda \rho}A_{\alpha \rho}}{u_\xi u^\xi }\bigr ) \sqrt{u_\delta u^\delta \det (g_{\epsilon \phi })}{u_\xi u^\xi}  \Bigr ]
		(u_\tau u^\tau +1) \nonumber \\
		&&=f(x^\mu ) (u_\tau u^\tau +1) \label{t3_2_eq2} \\
		\nonumber
		&&=0\;. \label{t3_2_eq3}
\eea
In the second equality, the total derivative of the first term vanishes due to the surface integral at infinity and \eq{t3_2_eq2} is absorbed into \eq{Sone} by redefining the Lagrange multiplier as, $\lambda'=\lambda +f(x^\mu ) $. From now on, we effectively disregard the covariant derivative $D_\lambda \sqrt{-\det (\hat{g}_{\rho \sigma})}$ by redefining $\lambda$.\\
With \eq{t3_eq3} and \eq{t3_2_eq3} the third term becomes
\bea
-\fr{1}{a+1}u^\alpha D_\lambda (\hat{g}^{\lambda \rho}A_{\alpha \rho})
			&=&-\fr{1}{a+1}(D_\lambda u_\beta)(\hat{g}^{\beta \alpha }-\fr{u^\beta u^\alpha}{a+1})\hat{g}^{\lambda \rho}A_{\alpha \rho} \nonumber \\
			&=&-\fr{1}{a+1}\Bigl [ -\fr{1}{2}(D_\lambda u_\beta -D_\beta u_\lambda )\hat{g}^{\beta\alpha}\hat{g}^{\lambda\rho}A_{\alpha \rho}\Bigr ]\nonumber \\
			&=&+\fr{1}{2}\fr{\hat{g}^{\alpha\beta}\hat{g}^{\rho \sigma}}{a+1}A_{\alpha \rho}A_{\lambda \beta} \nonumber \\
			&=&-\fr{1}{2}\fr{\hat{g}^{\alpha\beta}\hat{g}^{\rho \sigma}}{a+1}A_{\alpha \rho}A_{\beta \lambda}\label{t3}\;.
\eea
   Then let us look at the fourth term in \eq{a_Stotal01}
\bea
	&&\fr{1}{a+1}\hat{g}^{\alpha \beta} 
			\Bigl [ -D_\rho (u^\rho D_\alpha u_\beta )+D_\beta (u^\rho D_\alpha u_\rho )-\fr{u^\rho u^\sigma}{a+1} (D_\alpha u_\rho )D_\beta u_\sigma 
					+\fr{1}{2} \bigl \{ (D_\lambda u_\alpha )D^\lambda u_\beta -(D_\alpha u_\lambda )D_\beta u^\lambda \bigr \} \Bigr ]  \nonumber \\
	&&=\fr{1}{a+1}\Bigl [
		-\hat{g}^{\alpha \beta} D_\rho (u^\rho D_\alpha u_\beta )
		+\fr{1}{2} \left ( g^{\alpha \beta} +\fr{u^\alpha u^\beta}{a+1} \right )\bigl \{ (D_\lambda u_\alpha )D^\lambda u_\beta -(D_\alpha u_\lambda )D_\beta u^\lambda \bigr \}  \Bigr ] \nonumber \\
	&&=	\fr{1}{a+1}\Bigl [
		-\hat{g}^{\alpha \beta} D_\rho (u^\rho D_\alpha u_\beta ) 
		+\fr{1}{2} \left ( \fr{u^\alpha u^\beta}{a+1} \right )\bigl \{ -(D_\alpha u_\lambda )D_\beta u^\lambda \bigr \}  \Bigr ]\;,	
\eea
where the second term vanishes $D_\beta (u^\rho D_\alpha u_\rho )=0$ due to \eq{eq:uDu}, similarly the third term $\fr{u^\rho }{a+1} (D_\alpha u_\rho )u^\sigma D_\beta u_\sigma =0$ and $(u^\rho D_\rho u^\alpha )u^\beta D_\alpha u_\beta=0$ through the redefinition of the Lagrange multiplier $\lambda$   
\bea
\nonumber
	&&\fr{1}{a+1}\hat{g}^{\alpha \beta} 
			\Bigl [ -D_\rho (u^\rho D_\alpha u_\beta )+D_\beta (u^\rho D_\alpha u_\rho )-\fr{u^\rho u^\sigma}{a+1} (D_\alpha u_\rho )D_\beta u_\sigma 
					+\fr{1}{2} \bigl \{ (D_\lambda u_\alpha )D^\lambda u_\beta -(D_\alpha u_\lambda )D_\beta u^\lambda \bigr \} \Bigr ] \;, \nonumber \\
		&&=	\fr{1}{a+1}\Bigl [
		-\Bigl ( D_\rho (\hat{g}^{\alpha \beta} u^\rho D_\alpha u_\beta )
				-D_\rho (g^{\alpha \beta}+\fr{u^\alpha u^\beta}{a+1})u^\rho D_\alpha u_\beta \Bigr )
		-\fr{1}{2} \fr{1}{a+1}a_\lambda a^\lambda \bigr )  \Bigr ] \label{t4_2_eq1}\\
		&&=\fr{1}{a+1}\Bigl [
		\fr{1}{a+1}u^\rho (D_\rho u^\alpha )  u^\beta D_\alpha u_\beta + \fr{1}{a+1}u^\rho (D_\rho u^\beta ) u^\alpha D_\alpha u_\beta 
		-\fr{1}{2} \fr{1}{a+1}a_\lambda a^\lambda \bigr )  \Bigr ] \label{app:t4_2_eq3}\\
		&&=\fr{1}{2}\fr{1}{(a+1)^2}a_\lambda a^\lambda \label{app:t4}\;,
\eea
where the first term in \eq{t4_2_eq1} vanishes because the integrand is a  total derivative,  the first term in \eq{app:t4_2_eq3} can be ignored again by the redefinition of $\lambda$ and $a^\lambda :=u^\alpha D_\alpha u^\lambda$. Substituting \eq{t3} and \eq{app:t4} into \eq{a_Stotal01}, leads to
\bea
	S
	&=&\int \fr{c^4}{16\pi G}d^4x\tau d\sigma \sqrt{(g^{\xi \varepsilon }u_\xi u_\varepsilon  )\det (g_{\eta \iota} )}\Bigl [ \hat{g}^{\alpha \beta}R_{\alpha \beta} 
	  +\fr{1}{4}\fr{\hat{g}^{\alpha\beta}\hat{g}^{\rho \lambda}}{a+1}A_{\alpha \rho}A_{\beta \lambda}
		+\left (-\fr{1}{2}\fr{\hat{g}^{\alpha\beta}\hat{g}^{\rho \lambda}}{a+1}A_{\alpha \rho}A_{\beta \lambda} \right ) +\fr{1}{2}\fr{1}{(a+1)^2}a_\lambda a^\lambda  \nonumber \\
	&&	+\frac{16\pi G}{c^4} \lambda '(g^{\alpha \beta}u_\alpha u_\beta +1) \Bigr ]\nonumber \\
	&=&\int \fr{c^4}{16\pi G}d^4x\tau d\sigma \sqrt{(g^{\xi \varepsilon }u_\xi u_\varepsilon  )\det (g_{\eta \iota} )}\Bigl [ \hat{g}^{\alpha \beta}R_{\alpha \beta} 
	  -\fr{1}{4}\fr{\hat{g}^{\alpha\beta}\hat{g}^{\rho \lambda}}{a+1}A_{\alpha \rho}A_{\beta \lambda}
		+\fr{1}{2}\fr{1}{(a+1)^2}a_\lambda a^\lambda \nonumber \\
	&&	+\frac{16\pi G}{c^4} \lambda '(g^{\alpha \beta}u_\alpha u_\beta +1) \Bigr ]\;.\label{a_Stotal02}
\eea
Note that the second term in \eq{a_Stotal02} becomes
\bea
	-\fr{1}{4}\fr{\hat{g}^{\alpha\beta}\hat{g}^{\rho \lambda}}{a+1}A_{\alpha \rho}A_{\beta \lambda}
		&=&-\fr{1}{4}\fr{1}{a+1}\Bigl ( A_{\alpha \rho}A^{\alpha \rho}+\fr{2}{a+1}A^\beta _{~\rho}u^\rho A_{\beta \lambda}u^\lambda \Bigr )\;\nonumber \\
		&=&-\fr{1}{4}\fr{1}{a+1}A_{\alpha\beta}A^{\alpha\beta} -\fr{1}{2}\fr{1}{(a+1)^2}a_\beta a^\beta \label{app:AAaa},
\eea
where $(D^\beta u^\rho -D^\rho u^\beta )u_\rho=-u_\rho D^\rho u^\beta =-a^\beta$ due to the redefinition of Lagrange multiplier $\lambda$.\\
Finally \eq{a_Stotal02} is simplified to
\bea
		S
	&=&\int \fr{c^4}{16\pi G}d^4x\tau d\sigma \sqrt{(g^{\xi \varepsilon }u_\xi u_\varepsilon  )\det (g_{\eta \iota} )}\Bigl [ \hat{g}^{\alpha \beta}R_{\alpha \beta} 
	  -\fr{1}{4}\fr{g^{\alpha\beta}g^{\rho \lambda}}{a+1}A_{\alpha \rho}A_{\beta \lambda}
		+\frac{16\pi G}{c^4} \lambda '(g^{\alpha \beta}u_\alpha u_\beta +1) \Bigr ]\;.
\eea 
\section{Derivation of Spherically symmetric solution}\label{app:dR}
\subsection{Ricci tensor, Maxwell-like field strength and $P_{\mu\nu}$}
Under spherical symmetry
\bea
	ds^2_4=-B(t,r)c^2dt^2+A(t,r)dr^2+r^2\sin ^2\theta +r^2\sin^2d\phi ^2\;,
\eea
let us compute each component of \eq{Einstein}  
\bea
&G_{\mu\nu}^{\rm eff}:=R_{\mu \nu}-\frac{1}{2}g_{\mu \nu}R_{\alpha\beta}\hat{g}^{\alpha\beta}
	+\frac{1}{a+1}(R_{\nu\alpha}u_\mu+R_{\mu\alpha}u_\nu)u^{\alpha}
	-\frac{1}{2(a+1)}(g^{\alpha \beta}A_{\alpha\mu}A_{\beta\nu}
			-\frac{1}{4}g_{\mu \nu}A_{\alpha\beta}A^{\alpha\beta})\nonumber \\
&-\fr{1}{2(a+1)}D_\sigma D_\rho \Bigl [ \delta ^\rho _{~\nu}u_\mu u^\sigma +\delta ^\rho _{~\mu}u_\nu u^\sigma  -g^{\sigma \rho}u_\mu u_\nu- u^\rho u^\sigma g_{\mu\nu} \Bigr ] +\frac{16\pi G}{c^4}\lambda  u_\mu u_\nu=0. \nonumber
\eea

The only non-zero Christoffel symbols are
\bea
	\begin{array}{ccc}
			\Gamma ^t _{tt}=\fr{\dot{B}}{2B} &\Gamma ^t_{tr}=\fr{B'}{2B} &\Gamma ^t _{rr}=\fr{\dot{A}}{2B} \\
			\Gamma ^r _{tt}=\fr{B'}{2A} &\Gamma ^r _{tr} =\fr{\dot{A}}{2A} &\Gamma ^r _{rr}=\fr{A'}{2A} \\
			\Gamma ^r _{\theta\theta}=-\fr{r}{A} &\Gamma ^r_{\phi\phi}=-\fr{r\sin ^2\theta}{A} &	\Gamma ^\theta _{\theta r}=\fr{1}{r} \\
			\Gamma^\theta _{\phi\phi}=-\sin \theta\cos \theta &	\Gamma ^\phi _{\phi r}=\fr{1}{r} &\Gamma ^\phi _{\phi\theta}=\fr{\cos \theta}{\sin \theta}  
	\end{array}
\eea

From \eq{SPmetric}, as in \cite{WeinbergGandC} we can obtain the Ricci tensor by straightforward calculation
\bea
	R_{tt}&=&\fr{B''}{2A}-\fr{1}{4}\fr{A'B'}{A^2}-\fr{\ddot{A}}{2A}+\fr{1}{4}\fr{\dot{A}}{A}\fr{\dot{B}}{B}+\fr{1}{r}\fr{B'}{A}-\fr{1}{4}\fr{B'}{A}\fr{B'}{B}+\fr{1}{4}\fr{\dot{A}^2}{A^2}\;, \label{eq:Rictt}\\
	R_{tr}&=&\fr{1}{r}\fr{\dot{A}}{A}\;,\label{eq:Rictr}\\
	R_{rr}&=&\fr{\ddot{A}}{2B}-\fr{1}{4}\fr{\dot{A}}{B}\fr{\dot{B}}{B}-\fr{1}{2}\fr{B''}{B}+\fr{1}{4}\left (\fr{B'}{B}\right )^2+\fr{1}{4}\fr{A'}{A}\fr{B'}{B}+\fr{1}{r}\fr{A'}{A}-\fr{1}{4}\fr{\dot{A}}{A}\fr{\dot{A}}{B}\;, \label{eq:Ricrr}\\
	R_{\theta\theta}&=&1-\fr{1}{A}+\fr{r}{2}\fr{A'}{A^2}-\fr{r}{2}\fr{B'}{AB}\;,\label{eq:Ricthth}\\
	R_{\phi\phi}&=&\left (1-\fr{1}{A}+\fr{r}{2}\fr{A'}{A^2}-\fr{r}{2}\fr{B'}{AB}\right ) \sin ^2 \theta\;,\label{eq:Ricphph}
\eea
where $'$ and $\dot{}$ denote the derivative with respect to $r$ and $t$, respectively. 
The scalar curvature $R :=R_{\alpha \beta}g^{\alpha\beta}$ becomes
\bea
	R=-\fr{B''}{AB}+\fr{1}{2}\fr{A'B'}{A^2B}+\fr{\ddot{A}}{AB}-\fr{1}{2}\fr{\dot{A}\dot{B}}{AB^2}-\fr{2}{r}\fr{B'}{AB}+\fr{2}{r}\fr{A'}{A^2}+\fr{1}{2}\fr{(B')^2}{AB^2}-\fr{1}{2}\fr{\dot{A}^2}{A^2B}+\fr{2}{r^2}-\fr{2}{r^2}\fr{1}{A}\;.
\eea

Let us see some contributions from the $tt$-component of the Maxwell-like tensor $-\frac{1}{2(a+1)}(g^{\alpha \beta}A_{\alpha\mu}A_{\beta\nu}
			-\frac{1}{4}g_{\mu \nu}A_{\alpha\beta}A^{\alpha\beta})$. Noting 
\bea
	u_\mu =(-\sqrt{B},0,0,0)\;, \label{eq:u}
\eea
 and $A_{rt}=\pa _r u_t-\pa _tu_r=\pa_ru_t$ (other components are vanishing ), we obtain
\bea
	&&A^\alpha_{~t}A_{\alpha t}=g^{rr}A_{rt}A_{rt}=\fr{1}{4}\fr{(B')^2}{AB} \\
	&&A_{\alpha \beta}A^{\alpha\beta}=g^{\alpha\beta}(g^{tt}A_{\alpha t}A_{\beta t}+g^{rr}A_{\alpha t}A_{\beta r})=((\pa _r u_t )A_{rt}-(\pa _r u_t)A_{tr})=-\fr{1}{2}\fr{(B')^2}{AB^2} \\
	&&A^\alpha _{~t}A_{\alpha t}-\fr{1}{4} g_{tt}A_{\alpha \beta}A^{\alpha \beta}
		=\fr{1}{8}\fr{(B')^2}{AB}\;. 
\eea	
Then the $tt-$component of the Maxwell-like stress tensor term is	
\bea	
	&&-\frac{1}{2(a+1)}(g^{\alpha \beta}A_{\alpha t}A_{\beta t} 
			-\frac{1}{4}g_{tt}A_{\alpha\beta}A^{\alpha\beta})
		=-\fr{1}{16(a+1)}\fr{(B')^2}{AB}\;.\label{eq:AAtt}
\eea	
Similarly the other components are obtained as
\bea
 &&-\frac{1}{2(a+1)}(g^{\alpha \beta}A_{\alpha t}A_{\beta r}
			-\frac{1}{4}g_{t r}A_{\alpha\beta}A^{\alpha\beta})=0 \;,\label{eq:AAtr}\\
 &&-\frac{1}{2(a+1)}(g^{\alpha \beta}A_{\alpha r}A_{\beta r}
			-\frac{1}{4}g_{rr}A_{\alpha\beta}A^{\alpha\beta})
		=\fr{1}{16(a+1)}\fr{(B')^2}{B^2}	\;,	\label{eq:AArr}\\
 &&-\frac{1}{2(a+1)}(g^{\alpha \beta}A_{\alpha \theta}A_{\beta \theta}
			-\frac{1}{4}g_{\theta\theta}A_{\alpha\beta}A^{\alpha\beta})
		=\fr{r^2}{16(a+1)}\fr{(B')^2}{AB^2}	\;,\label{eq:AAthth}\\
 &&-\frac{1}{2(a+1)}(g^{\alpha \beta}A_{\alpha \phi}A_{\beta \phi}
			-\frac{1}{4}g_{\phi\phi}A_{\alpha\beta}A^{\alpha\beta})
		=\fr{r^2\sin ^2 \theta}{16(a+1)}\fr{(B')^2}{AB^2}	\;.\label{eq:AAphph}		
\eea 

Let us turn to the derivation of $\hat{g}^{\alpha\beta}\delta R_{\alpha\beta}=g^{\alpha\beta} \delta R_{\alpha\beta}+\fr{1}{a+1}u^\alpha u^\beta \delta R_{\alpha \beta}$. The variation of the Ricci tensor $\delta R_{\mu\nu}$ is written as 
\bea
 \delta R^{\alpha}_{~\mu\alpha\nu}
	&=&\pa _{\alpha}\delta \Gamma ^{\alpha}_{\mu\nu} 
	-\pa_\nu \delta \Gamma ^{\alpha}_{\mu \alpha}
	+\delta (\Gamma^{\alpha}_{\alpha\beta} 
     \Gamma^{\beta}_{\nu\mu})
    -\delta (\Gamma^{\alpha}_{\nu \beta} 
     \Gamma^{\beta}_{\alpha\mu} ) \\
     \nonumber
	&=& D_\alpha \delta \Gamma ^\alpha _{\mu\nu}
		-D_\nu \delta \Gamma^\alpha _{\mu\alpha}\;.
\eea
Then the term $g^{\alpha \beta}\delta R_{\alpha \beta}$ can be made a total derivative form
\bea
	g^{\mu\nu}\delta R_{\mu\nu}=g^{\mu \nu}
		\bigl (
			 D_\alpha (g^{\mu\nu} \delta \Gamma ^\alpha _{\mu\nu})
		-D_\nu (g^{\mu\nu} \delta \Gamma^\alpha _{\mu\alpha})
		\bigr )\label{eq:gdR}\;,
\eea
which vanishes by an appropriate boundary condition at infinity in the action integral, while the term $\fr{u^\alpha u^
\beta}{a+1}\delta R_{\alpha \beta}$ cannot be integrated out. Therefore, when we derive field equations from the action (\ref{action}) the term $u^\alpha u^\beta \delta R_{\alpha\beta}$ remains non-zero\;, 
\bea
	u^\mu u^\nu \delta R_{\mu\nu}
		&=&u^\mu u^\nu D_\alpha \delta \Gamma ^\alpha _{\mu\nu}
		-u^\mu u^\nu D_\nu \delta \Gamma^\alpha _{\mu\alpha} \nonumber \\
		&=&-D_\alpha (u^\mu u^\nu )\delta \Gamma^\alpha _{\mu\nu}
		+D_\nu (u^\mu u^\nu)\delta \Gamma ^\alpha _{\mu\alpha} \;.\label{app:dR_1}
\eea
Noting that
\bea
	\pa _\nu \delta g_{\mu\rho}+\pa _\mu \delta g_{\nu\rho}-\pa _\rho \delta g_{\mu\nu}
	=D_\nu \delta g_{\mu\rho}+D_\nu \delta g_{\nu\rho}-D_\rho\delta g_{\mu\nu}+2\Gamma ^\tau _{\mu\nu}\delta g_{\tau \rho},
\eea  
we see that $\delta \Gamma ^\alpha _{\mu\nu}$ becomes
\bea
 \delta \Gamma ^\alpha _{\mu\nu}
	&=&\fr{1}{2}\delta g^{\alpha \rho}(g_{\mu\rho,\nu}+g_{\nu\rho,\mu}-g_{\mu\nu,\rho})+\fr{1}{2}g^{\alpha \rho}(\delta g_{\mu\rho,\nu}+\delta g_{\nu\rho,\mu}-\delta g_{\mu\nu,\rho}) \nonumber \\
	&=&\fr{1}{2}g^{\alpha\rho}(D_\nu \delta g_{\mu\rho}+D_\mu \delta g_{\nu\rho}-D_\rho\delta g_{\mu\nu})\label{app:dGam_1}.
\eea
From \eq{app:dR_1} and \eq{app:dGam_1}, the additional term becomes 
\bea
	\fr{1}{a+1}u^\mu u^\nu \delta R_{\mu\nu}
	&=&\fr{1}{2}\fr{1}{a+1}\Bigl [ 
		-D^\rho (u^\mu u^\nu )D_\nu\delta g_{\mu \rho}
		-D^\rho (u^\mu u^\nu )D_\mu \delta g_{\nu\rho}
		+D^\rho (u^\mu u^\nu )D_\rho \delta g_{\mu\nu}
		+D_\nu (u^\mu u^\nu) g^{\alpha \rho}D_\mu \delta g_{\alpha\rho}
	\Bigr ]\;. \nonumber \\
	\nonumber \\
	&=&-\fr{1}{2(a+1)}D_\sigma D_\rho \Bigl [ \delta ^\rho _{~\nu}u_\mu u^\sigma +\delta ^\rho _{~\mu}u_\nu u^\sigma-g^{\sigma \rho}u_\mu u_\nu-u^\lambda u^\rho g_{\mu\nu} \Bigr ] \delta g^{\mu \nu} \;. \label{app:dl_uuR}
\eea
It is convenient to calculate the second order covariant derivative, $D_\mu D_\nu u_\rho$ and list up its components as
\bea
	\begin{array}{|c|c|c|c|c|}
	\hline
	D_\mu u_\nu			&u_t	&u_r	&u_\theta	&u_\phi\\
	\hline
	D_t 		&0	&\fr{\sqrt{B'}}{2\sqrt{B}}	&0	&0 \\
	\hline
	D_r 		&0	&\fr{\sqrt{\dot{A}}}{2\sqrt{B}}	&0	&0 \\
	\hline
	D_\theta &0	&0	&0	&0 \\
	\hline
	D_\phi &0	&0	&0	&0\\
	\hline	
	\end{array}\;.
\eea

For the $-2(a+1) P_{tt}$, we need to calculate $D_\sigma D_\rho \Bigl [ \delta ^\rho _{~t}u_t u^\sigma +\delta ^\rho _{~t}u_t u^\sigma-g^{\sigma \rho}u_t u_t-u^\lambda u^\rho g_{tt} \Bigr ]$, each term of which can be written as 
\bea
	2 g^{\rho\sigma} D_\sigma D_t(u_t u_\beta )
		&=&2 g^{\rho\sigma}\Bigl [	 u_\rho D_\sigma D_t u_t 
									+u_t D_\rho D_t u_\sigma \Bigr ] \nonumber \\
		&=&\bigl ( -\fr{1}{2}\fr{(B')^2}{AB}\bigr )
		 +\bigl (-\fr{B''}{A}+\fr{1}{2}\fr{(B')^2}{AB}+\fr{1}{2}\fr{\dot{A}^2}{A^2}
           +\fr{1}{2}\fr{A'B'}{A^2}-\fr{2}{r}\fr{B'}{A} \bigr ) \nonumber \\
		&=&	-\fr{B''}{A}
			+\fr{1}{2}\fr{\dot{A}^2}{A^2}
            +\fr{1}{2}\fr{A'B'}{A^2}
			-\fr{2}{r}\fr{B'}{A}\;,		\\			
	- g^{\rho\sigma} D_\rho D_\sigma (u_t u_t )
			&=&-2g^{\rho\sigma} D_\rho D_\sigma u_t 
			=\fr{1}{2}\fr{(B')^2}{AB}-\fr{1}{2}\fr{\dot{A}^2}{A^2}\;,
\eea
\bea
	&&- g_{tt}g^{\alpha\rho}g^{\beta \sigma}D_\alpha D_\beta (u_\rho u^\sigma )\nonumber \\
			&&=B g^{\alpha\rho}g^{\beta \sigma}
				\Bigl [
					u_\rho D\alpha D_\beta u_\sigma
					+(D_\alpha u_\rho )D_\beta u_\sigma 
					+(D_\alpha u_\sigma )D_\beta u_\rho
					+u_\sigma D_\alpha D_\beta u_\rho 
				\Bigr ] \\
			&&=\bigl ( \fr{1}{2}\fr{\ddot{A}}{A}
						-\fr{1}{4}\fr{\dot{A}\dot{B}}{AB}
						-\fr{1}{2}\fr{\dot{A}^2}{A^2}\bigr ) 
				+\bigl ( \fr{1}{4}\fr{\dot{A}^2}{A^2}\bigr )
				+\bigl ( \fr{1}{4}\fr{\dot{A}^2}{A^2}\bigr )
				+\bigl ( \fr{1}{2}\fr{B''}{A} 
						-\fr{1}{4}\fr{(B')^2}{AB}
						-\fr{1}{4}\fr{\dot{A}^2}{A^2}
						-\fr{1}{4}\fr{A'B'}{A^2}
						+\fr{1}{r}\fr{B'}{A}\bigr ) \nonumber \\
			&&= \fr{1}{2}\fr{\ddot{A}}{A}
						-\fr{1}{4}\fr{\dot{A}\dot{B}}{AB}
						-\fr{1}{4}\fr{\dot{A}^2}{A^2}
						+ \fr{1}{2}\fr{B''}{A} 
						-\fr{1}{4}\fr{(B')^2}{AB}
						-\fr{1}{4}\fr{A'B'}{A^2}
						+\fr{1}{r}\fr{B'}{A}\;.				
\eea
Putting them together we obtain 
\bea
	P_{tt}	&=&-\fr{1}{2(a+1)}\Bigl [ -\fr{1}{2}\fr{B''}{A}+\fr{1}{4}\fr{A'B'}{A^2}+\fr{1}{4}\fr{(B')^2}{AB}+\fr{1}{2}\fr{\ddot{A}}{A}-\fr{1}{4}\fr{\dot{A}\dot{B}}{AB}-\fr{1}{r}\fr{B'}{A}-\fr{1}{4}\fr{\dot{A}^2}{A^2}\Bigr ]\;. \label{eq:Ptt} 
\eea	
We also see that $P_{tr}=0$ since
\bea
	-2(a+1)P_{tr}
	&=&D_\sigma D_t (u_r u^\sigma) +D_\sigma D_r(u_t u^\sigma ) -g^{\sigma \rho}D_\sigma D_\rho (u_t u_r)  \nonumber \\
	&=&\bigl ( \fr{1}{2}\fr{\dot{B}'}{B}-\fr{1}{2}\fr{\dot{B}B'}{B^2}\bigr )
	   +\bigl ( -\fr{\dot{A}'}{2A}+\fr{1}{2}\fr{\dot{A}}{A}\fr{A'}{A}-\fr{1}{r}\fr{\dot{A}}{A} \bigr )\nonumber \\
	  && +\bigr (	-\fr{1}{2}\fr{\dot{B}'}{B}
				+\fr{1}{2}\fr{\dot{A}'}{A}
				+\fr{1}{2}\fr{\dot{B}B'}{B^2}
				-\fr{1}{4}\fr{\dot{A}}{A}\fr{A'}{A}
				+\fr{1}{r}\fr{\dot{A}}{A}
				-\fr{1}{4}\fr{\dot{A}}{A}\fr{A'}{A} \bigr ) \nonumber \\
	&=&0 \;. \label{eq:Ptr}
\eea

For $P_{rr}$, we see that 
\bea
	-2(a+1)P_{rr}
		&=&2D_\sigma D_r (u_r u^\sigma )-g^{\sigma \rho}D_\sigma D_\rho (u_r u_r)-D_\sigma D_\rho(u^\sigma u^\rho) g_{rr} \nonumber \\
		&=&\bigl ( \fr{\ddot{A}}{B}-\fr{1}{2}\fr{\dot{A}\dot{B}}{B^2}-\fr{1}{2}\fr{(B')^2}{B^2} \bigr )
			+\bigl ( \fr{1}{2}\fr{(B')^2}{B^2}-\fr{1}{2}\fr{\dot{A}^2}{AB} \bigr )\nonumber \\
		&&	+\bigl ( -\fr{1}{2}\fr{\ddot{A}}{B}
					+\fr{1}{4}\fr{\dot{A}\dot{B}}{B^2}
					+\fr{1}{4}\fr{\dot{A}^2}{AB}
					-\fr{1}{2}\fr{B''}{B}
					+\fr{1}{4}\fr{(B')^2}{B^2}
					+\fr{1}{4}\fr{A'B'}{AB}
					-\fr{1}{r}\fr{B'}{B}
			\bigr ) \nonumber \\
		&=&\fr{1}{2}\fr{\ddot{A}}{B}-\fr{1}{4}\fr{\dot{A}\dot{B}}{B^2}
			-\fr{1}{4}\fr{\dot{A}^2}{AB}-\fr{1}{2}\fr{B''}{B}
			+\fr{1}{4}\fr{(B')^2}{B^2}+\fr{1}{4}\fr{A'B'}{AB}
			-\fr{1}{r}\fr{B'}{B}\;.
\eea
Then,
\bea
	P_{rr}&=&-\fr{1}{2(a+1)}\Bigl [ \fr{1}{2}\fr{\ddot{A}}{B}-\fr{1}{4}\fr{\dot{A}}{A}\fr{\dot{B}}{B}-\fr{1}{4}\fr{\dot{A}^2}{AB}-\fr{1}{2}\fr{B''}{B}+\fr{1}{4}\left ( \fr{B'}{B}\right )^2+\fr{1}{4}\fr{A'}{A}\fr{B'}{B}-\fr{1}{r}\fr{B'}{B} \Bigr ]\;. \label{eq:Prr}
\eea
Similarly we see that
\bea
	-2(a+1) P_{\theta\theta}
	&=&2D_\sigma D_\theta (u_\theta u^\sigma ) -g^{\sigma \rho}D_\sigma D_\rho (u_\theta u_\theta)
		-  g_{\theta\theta} D_\sigma D_\rho (u^\sigma u^\rho)  \nonumber \\
	&=&(-r^2)
			\Bigl ( 
				\fr{1}{2}\fr{\ddot{A}}{AB}
				-\fr{1}{4}\fr{\dot{A}\dot{B}}{AB^2}
				-\fr{1}{4}\fr{\dot{A}^2}{A^2B}
				+\fr{1}{2}\fr{B''}{AB}
				-\fr{1}{4}\fr{(B')^2}{AB^2}
				-\fr{1}{4}\fr{A'B'}{A^2B}
				+\fr{1}{r}\fr{B'}{AB}
			\Bigr )\;,
\eea
so that
\bea
	P_{\theta\theta}&=&-\fr{-r^2}{2(a+1)}\Bigl [ \fr{1}{2}\fr{\ddot{A}}{AB}-\fr{1}{4}\fr{\dot{A}\dot{B}}{AB^2}-\fr{1}{4}\fr{\dot{A}^2}{A^2B}+\fr{1}{2}\fr{B''}{AB}-\fr{1}{4}\fr{(B')^2}{AB^2}-\fr{1}{4}\fr{A'B'}{A^2B}+\fr{1}{r}\fr{B'}{AB} \Bigr ]\;,\label{eq:Pthth} \\
	P_{\phi\phi}&=&-\fr{-r^2\sin ^2\theta}{2(a+1)}\Bigl [ \fr{1}{2}\fr{\ddot{A}}{AB}-\fr{1}{4}\fr{\dot{A}\dot{B}}{AB^2}-\fr{1}{4}\fr{\dot{A}^2}{A^2B}+\fr{1}{2}\fr{B''}{AB}-\fr{1}{4}\fr{(B')^2}{AB^2}-\fr{1}{4}\fr{A'B'}{A^2B}+\fr{1}{r}\fr{B'}{AB} \Bigr ] \;.\label{eq:Pphph}
\eea

\subsection{The Einstein equation in spherically symmetric case}
Substituting the components of the Ricci tensor, the 4-vector field components \eq{eq:u}, the Maxwell-like components \eq{eq:AAtt}, \eq{eq:AAtr}, \eq{eq:AArr}, \eq{eq:AAthth}, \eq{eq:AAphph}, the correction terms Eqs.~(\ref{eq:Ptt}), (\ref{eq:Ptr}), (\ref{eq:Prr}), (\ref{eq:Pthth}), (\ref{eq:Pphph}) into \eq{Einstein} again, we calculate the effective Einstein tensor $G_{\mu\nu}^{\rm eff}$ in the spherically symmetric 4-dimensional metric as
\bea
G_{\mu\nu}^{\rm eff}&:=&R_{\mu \nu}-\frac{1}{2}g_{\mu \nu}R_{\alpha\beta}\hat{g}^{\alpha\beta}
	+\frac{1}{a+1}(R_{\nu\alpha}u_\mu+R_{\mu\alpha}u_\nu)u^{\alpha}
	-\frac{1}{2(a+1)}(g^{\alpha \beta}A_{\alpha\mu}A_{\beta\nu}
			-\frac{1}{4}g_{\mu \nu}A_{\alpha\beta}A^{\alpha\beta})\nonumber \\
&&+P_{\mu\nu} +\frac{16\pi G}{c^4}\lambda  u_\mu u_\nu=0\;.\nonumber 
\eea

  Each component of the effective Einstein tensor \eq{Einstein} can be summarized as
\bea
	G_{tt}^{\rm eff}&=&\fr{B}{r^2}+\frac{16\pi G}{c^4} \lambda B-\fr{B}{r^2A}+\fr{BA'}{rA^2}-\fr{1}{a+1}\fr{B'}{r A}+\fr{1}{4(a+1)}\fr{A'}{A}\fr{B'}{A}+\fr{3}{16(a+1)}\fr{(B')^2}{AB}-\fr{1}{2}\fr{1}{a+1}\fr{B''}{A}\nonumber \\
	&&-\fr{1}{4}\fr{1}{a+1}\fr{\dot{A}^2}{A^2}-\fr{1}{4}\fr{1}{a+1}\fr{\dot{A}\dot{B}}{AB}+\fr{1}{2}\fr{1}{a+1}\fr{\ddot{A}}{A} \;,\label{eq:G11} \\
	G_{tr}^{\rm eff}&=&\fr{a}{a+1}\fr{\dot{A}}{rA}\;, \label{eq:G12}\\
	G_{rr}^{\rm eff}&=&\fr{1}{r^2}-\fr{A}{r^2}+\fr{1}{r}\fr{B'}{B}+\fr{1}{16}\fr{1}{a+1}\fr{(B')^2}{B^2}\;, \label{eq:G22} \\
	G_{\theta\theta}^{\rm eff}&=&r^2 \Bigl [-\fr{1}{2r}\fr{A'}{A^2}+\fr{1}{2r}\fr{B'}{AB}-\fr{1}{4}\fr{A'B'}{A^2B}-\fr{1}{4}\fr{(B')^2}{AB^2}-\fr{1}{16}\fr{1}{a+1}\fr{(B')^2}{AB^2}+\fr{1}{2}\fr{B''}{AB}\nonumber \\
	&&+\fr{1}{4}\fr{\dot{A}^2}{A^2B}-\fr{1}{4}\fr{1}{a+1}\fr{\dot{A}^2}{A^2B}+\fr{1}{4}\fr{\dot{A}\dot{B}}{AB^2}-\fr{1}{4}\fr{1}{a+1}\fr{\dot{A}\dot{B}}{AB^2}-\fr{1}{2}\fr{\ddot{A}}{AB}+\fr{1}{2}\fr{1}{a+1}\fr{\ddot{A}}{AB} \Bigr ]\;, \label{eq:G33}\\
	G_{\phi\phi}^{\rm eff}&=&r^2\sin^2 \theta \Bigl [-\fr{1}{2r}\fr{A'}{A^2}+\fr{1}{2r}\fr{B'}{AB}-\fr{1}{4}\fr{A'B'}{A^2B}-\fr{1}{4}\fr{(B')^2}{AB^2}-\fr{1}{16}\fr{1}{a+1}\fr{(B')^2}{AB^2}+\fr{1}{2}\fr{B''}{AB}\nonumber \\
	&&+\fr{1}{4}\fr{\dot{A}^2}{A^2B}-\fr{1}{4}\fr{1}{a+1}\fr{\dot{A}^2}{A^2B}+\fr{1}{4}\fr{\dot{A}\dot{B}}{AB^2}-\fr{1}{4}\fr{1}{a+1}\fr{\dot{A}\dot{B}}{AB^2}-\fr{1}{2}\fr{\ddot{A}}{AB}+\fr{1}{2}\fr{1}{a+1}\fr{\ddot{A}}{AB} \Bigr ]\;. \label{eq:G44}
\eea  
\subsubsection{Derivation of the spherically symmetric solution}\label{app:sph_sym_sol}
Let us obtain explicit expressions for $r,B,A$ and $\lambda$ as functions of $V$.
We can integrate $\fr{dV}{dr}=-\fr{V}{r}\fr{\alpha ^2 +\alpha ^2 V+V^2}{\alpha ^2}$to give the expression $r(V)$, noting a factorization 
\bea
	&&V^2 +\alpha ^2 V+\alpha ^2 =(V-a)(V-b)\;, 
\eea
we obtain
\bea	
	&&\int \fr{dV}{V(V- v_1)(V- v_2)}=-\fr{1}{\alpha ^2}\int \fr{dr}{r}\;, \label{eq:VVV}
\eea
where $v_1:=\fr{1}{2}(-\alpha ^2+\sqrt{\alpha ^4-4\alpha ^2})$ and $v_2:=\fr{1}{2}(-\alpha ^2-\sqrt{\alpha ^4-4\alpha ^2})$.
By partial fractional decomposition, the left hand side of the integrand of \eq{eq:VVV} becomes
\bea
	\fr{1}{V(V- v_1)(V- v_2)}
		&=&\fr{1}{ v_1- v_2} \left ( \fr{1}{V(V- v_1)}-\fr{1}{V(V- v_2)} \right ) \nonumber \\
		&=&\fr{1}{ v_1( v_1- v_2)}\fr{1}{V- v_1}+\fr{1}{ v_1 v_2}\fr{1}{V}-\fr{1}{ v_2( v_1- v_2)}\fr{1}{V- v_2}\;.
\eea
Then, the left hand side of \eq{eq:VVV} can be integrated to give
\bea
	\int \fr{dV}{V(V- v_1)(V- v_2)}
		&=&\fr{1}{ v_1 v_2} \log \fr{(V- v_1)^{\fr{ v_2}{ v_1- v_2}}V}{(V- v_2)^{\fr{ v_1}{ v_1- v_2}}} 
		=\fr{1}{\alpha ^2}\log \fr{(V- v_1)^{-\fr{\varepsilon  +1}{2}}V}{(V- v_2)^{-\fr{\varepsilon -1}{2}}} \nonumber \\
		&=&\fr{(V-v_2)^\fr{\varepsilon -1}{2}V}{(V-v_1)^\fr{\varepsilon  +1}{2}}\;,
\eea

where $ v_1- v_2=\sqrt{\alpha ^4 -4\alpha ^2}$,
 $\varepsilon  :=\fr{\alpha ^2}{\sqrt{\alpha ^4-4\alpha ^2}}\geq 1$, $\fr{ v_2}{ v_1- v_2}=\fr{1}{2}\left ( \fr{-\alpha ^2}{\sqrt{\alpha ^4-4\alpha ^2}}-1\right )=-\fr{\varepsilon  +1}{2}$, 
$\fr{v_1}{v_1- v_2}=\fr{1}{2}\left ( \fr{-\alpha ^2}{\sqrt{\alpha ^4-4\alpha ^2}}+1\right )=-\fr{\varepsilon -1}{2}$ and $\fr{1}{v_1 v_2}=\fr{1}{\alpha ^2}$.
The \eq{eq:VVV} becomes
\bea
	&&\fr{(V-v_2)^\fr{\varepsilon -1}{2}V}{(V- v_1)^\fr{\varepsilon  +1}{2}}
		=\fr{1}{\alpha ^2}\log (r^{-1} C) 
\eea
Or setting $C=1$, we obtain
\bea
	r(V)=\fr{(V-v_1)^{\fr{\varepsilon  +1}{2}}}{(V- v_2)^{\fr{\varepsilon -1}{2}}V}\;.\label{eq:r_V_app}
\eea
  
As $g_{tt}(V)=B(V)$ is already concretely given in the main text as in \eq{eq:F_V} -- the \eq{eq:Breg_V} and so as $g_{rr}(V)=A(V)$, we will not repeat them here. From \eq{eq:A_B} we need $\fr{dB}{dr}$.
\bea
	\fr{dB}{dr}
		&=&\fr{dB}{dV}\fr{dV}{dr} \nonumber \\
		&=&\varepsilon  \left (\fr{V-v_2}{V-v_1} \right ) ^{-1}B(V)\fr{-(v_1- v_2)}{(V- v_1)^2}\left (-\fr{V}{r} \right ) \fr{(V-v_1)(V-v_2)}{\alpha ^2} \nonumber \\
		&=&\varepsilon  \fr{v_1- v_2}{\alpha ^2}\fr{V}{r}B \nonumber \\
		&=&\fr{V}{r}B(V)\;. \label{eq:dB}
\eea
Substitution of \eq{eq:dB} into \eq{eq:A_B} leads to an expression for $A$,
\bea
	A
		&=&1+r \left ( \fr{V}{r} \right )+\fr{r^2}{\alpha ^2} \left (\fr{V}{r}\right )^2 \nonumber \\
		&=&1+V+\fr{V^2}{\alpha ^2}\;.\label{app:A_V}
\eea

Now let us have a closer look at the Lagrange multiplier $\lambda (V)$. The $tt-$component of the effective Einstein equation $G_{tt}^{\rm eff}=0$ (see \eq{eq:G11}) on the basis of the Birkhoff theorem proven in \eq{eq:dotA}
\bea
	&&\lambda =-\fr{c^4}{\pi G A^2 \alpha^2 } \Bigl [-\fr{1}{2}A\fr{B''}{B}+\fr{1}{4}A'\fr{B'}{B}-\fr{1}{r}A\fr{B'}{B}\Bigr ] 
	-\fr{c^4}{16\pi G}\left ( \fr{1}{r}\fr{A'}{A^2}+\fr{1}{r^2}-\fr{1}{r^2}\fr{1}{A} \right )-\fr{c^4}{\pi GA^2\alpha ^2}\fr{3}{16}A\left (\fr{B'}{B} \right )^2\;,  \label{eq:l_AB}
\eea
where $'$ denotes the derivative with respect to $r$, and  $\alpha ^2 :=16(1+a)$ .
Noting that using \eq{eq:dV}, \eq{eq:A_V}, \eq{eq:dB} and $\fr{dV}{dr}=-\fr{V}{r}A$, we obtain,
\bea
	\fr{d^2B}{dr^2}
		&=&\fr{d}{dr} \left ( \fr{V}{r}B(V) \right )
		=\fr{1}{r}\fr{dV}{dr}B-\fr{V}{r^2}B+\fr{V}{r}\fr{dB}{dr} \nonumber \\
		&=&\fr{V}{r^2}B\left (-\fr{V^2}{\alpha ^2}-2 \right )\;, \label{eq:ddB} \\
	\fr{dA}{dr}	
		&=&\fr{dA}{dV}\fr{dV}{dr}\nonumber \\
		&=&(-\fr{V}{r}-\fr{2V^2}{r\alpha ^2} )A\;, \label{eq:dA}
\eea
and then by substituting $\fr{B'}{B}=\fr{V}{r}, \fr{B''}{B}=-\fr{2V}{r^2}-\fr{V^3}{\alpha ^2 r^2}, A=\fr{V^2 +\alpha ^2 V +\alpha ^2}{\alpha ^2}$ and $A'= (-\fr{V}{r}-\fr{2V^2}{r\alpha ^2} )A$, into \eq{eq:l_AB} we obtain $\lambda $ as
\bea
	\lambda 
		&=& -\fr{c^4}{\pi G A \alpha ^2}\left ( -\fr{1}{4}\fr{V^2}{r^2}+\fr{3}{16}\fr{V^2}{r^2}\right ) -\fr{c^4}{\pi G A} \left ( \fr{1}{16r} (-\fr{V}{r}-\fr{2V^2}{r\alpha ^2}) +\fr{A}{16r^2}-\fr{1}{16}\fr{1}{r^2} \right ) \nonumber \\
		&=&\fr{c^4}{8\pi GA}\fr{V^2}{r^2 \alpha ^2}\;. \label{eq:l_V}
\eea
\subsubsection{Checking the equation of motion for the vector field $u^\mu(x)$}\label{app:check_u}

In the previous subsection \ref{app:sph_sym_sol}, we have obtained the analytic functions $r(V),B(V),A(V),\lambda (V)$. By using the equation of motion for the field $u^\mu$
\bea
 \fr{1}{a+1}\bigl [ D_\alpha A^{\alpha \mu} +2R^{\mu }_{~\alpha}u^\alpha \Bigr ] +\frac{32\pi G}{c^4}\lambda u^\mu =0 \label{eq:ueom}
\eea
we check if the analytical functions are indeed exact solutions of a set of whole equation of motion. Now we look at the $t-$component of \eq{eq:ueom}. We list the explicit expression for the covariant derivative $D_\mu A_{\nu \rho}$. Noting every component $\mu, \nu, \rho$ runs from $t$ to $\phi$. 
\bea
		\begin{array}{|c|c|c|c|c|}
		\hline
			D_\mu A_{\nu \rho}	&\nu =t&\nu =r&\nu =\theta & \nu=\phi \\
		\hline
		\mu =t 		&\bc 0 \\ \fr{\dot{B}'}{2\sqrt{B}}-\fr{\dot{B}B'}{2B^\fr{3}{2}}-\fr{1}{4}\fr{\dot{A}B'}{AB^\fr{1}{2}}\\0\\0\ec
							&\bc -D_tA_{tr} \\ 0\\0\\0\ec 
							&\bc 0 \\ 0\\0\\0\ec
							&\bc 0 \\ 0\\0\\0\ec \\
		\hline
		\mu =r 		&\bc 0 \\ -D_rA_{rt}\\0\\0\ec
							&\bc -\fr{B''}{2B^\fr{1}{2}}+\fr{1}{2}\fr{(B')^2}{B^\fr{3}{2}}+\fr{1}{4}\fr{A'B'}{AB^\fr{1}{2}} \\ 0\\0\\0\ec 
							&\bc 0 \\ 0\\0\\0\ec
							&\bc 0 \\ 0\\0\\0\ec \\
		\hline
		\mu =\theta &\bc 0 \\ 0\\ \fr{r}{2A}\fr{B'}{B^\fr{1}{2}} \\0\ec
							&\bc 0 \\ 0\\ 0 \\0\ec 
							&\bc -D_\theta A_{t\theta} \\ 0\\0\\0\ec
							&\bc 0 \\ 0\\0\\0\ec \\
		\hline
		\mu =\phi	&\bc 0 \\ 0\\ 0 \\ \fr{r}{2A}\fr{B'}{B^\fr{1}{2}}\sin ^2\theta \ec
							&\bc 0 \\ 0\\ 0 \\0\ec 
							&\bc 0 \\ 0\\ 0\\0\ec
							&\bc -D_tA_{\phi t} \\ 0\\0\\0\ec \\
		\hline	
		\end{array}\label{tb:DA}
\eea

Now the $t-$component of the field equation read as
\bea
 \fr{1}{a+1}\bigl [ D_\alpha A^{\alpha t} +2R^{t }_{~\alpha}u^\alpha \Bigr ] +\frac{32\pi G}{c^4}\lambda u^t =0\;. \label{eq:ueom_t}
\eea

The first term in \eq{eq:ueom_t} is
\bea
	\fr{1}{a+1}g^{t\alpha}g^{\beta\rho}D_\rho A_{\beta \alpha}
		&=&\fr{1}{a+1} \left ( \fr{B''}{2AB^\fr{3}{2}}-\fr{1}{2}\fr{(B')^2}{AB^\fr{5}{2}}-\fr{1}{4}\fr{A'B'}{A^2B^\fr{3}{2}}+\fr{1}{r}\fr{B'}{AB^\fr{3}{2}} \right )\;
\eea
and the second term is
\bea
	\fr{2}{a+1}g^{t\alpha}R_{\alpha \rho}u^\rho
		&=&\fr{1}{a+1} \left ( -\fr{B''}{AB^\fr{3}{2}}+\fr{1}{2}\fr{A'B'}{A^2B^\fr{3}{2}}+\fr{\ddot{A}}{AB^\fr{3}{2}}-\fr{1}{2}\fr{\dot{A}\dot{B}}{AB^\fr{5}{2}}-\fr{2}{r}\fr{B'}{AB^\fr{3}{2}}+\fr{1}{2}\fr{(B')^2}{AB^\fr{5}{2}}-\fr{1}{2}\fr{\dot{A}^2}{A^2B^\fr{3}{2}} \right )\;,
\eea
where \eq{eq:Rictt} is used. Writing $\alpha^2 :=16(a+1)$ the field equation of $u^t$ \eq{eq:ueom_t} reduces to
\bea
	&&\fr{16}{A\sqrt{B}}\Bigl [ \fr{1}{\alpha ^2} \bigl ( -\fr{1}{2}\fr{B''}{B}+\fr{1}{4}\fr{A'}{A}\fr{B'}{B}-\fr{1}{r}\fr{B'}{B}+\fr{\ddot{A}}{B}-\fr{1}{2}\fr{\dot{A}\dot{B}}{B^2}-\fr{1}{2}\fr{\dot{A}^2}{AB}\bigr )+\frac{2\pi G}{c^4}\lambda A \Bigr ]=0\;. \label{eq:ueom_t_red}
\eea
If we substitute our solution $B(V),A(V),\lambda(V)$, with (\ref{eq:dB}), (\ref{app:A_V}), (\ref{eq:ddB}), (\ref{eq:dA}), $\dot{B}=0$, Eq.(\ref{eq:dotA}) leads to
\bea	
	&&\fr{16}{A\sqrt{B}}\Bigl [ \fr{1}{\alpha ^2} \bigl ( -\fr{V^2}{4r^2} \bigr )+\frac{2\pi G}{c^4}\lambda A \Bigr ]=0\;. \label{eq:ueom_t2}
\eea

Now turn to the expression for $\lambda$. From \eq{eq:G11}, $G_{tt}^{\rm eff}=0$ with $\alpha ^2=16(a+1)$ we have
\bea
	\frac{2\pi G}{c^4}\lambda A=\fr{1}{\alpha^2}\bigl ( \fr{B''}{B}-\fr{1}{2}\fr{A'B'}{AB}+\fr{2}{r}\fr{B'}{B}-\fr{\ddot{A}}{B}+\fr{1}{2}\fr{\dot{A}\dot{B}}{B^2}+\fr{1}{2}\fr{\dot{A}^2}{AB} \bigr )-\fr{3}{8\alpha ^2}\fr{(B')^2}{B^2}-\fr{1}{8r}\fr{A'}{A}-\fr{A}{8r^2}+\fr{1}{8r^2}\;.\label{eq:lA}
\eea
If we substitute our solution $B(V),A(V),\lambda(V)$, with Eqs.~(\ref{eq:dB})-(\ref{eq:dA}), $\dot{B}=0$ and $\dot{A}=0$ we see that 
\bea
	\frac{2\pi \lambda G}{c^4}A =\fr{1}{4}\fr{V^2}{\alpha ^2r^2}. 
\eea 

By inserting the \eq{eq:lA} into the \eq{eq:ueom_t} the $t-$ component of the field equation becomes
\bea
	\fr{16}{\sqrt{B}} \Bigl [ \fr{1}{\alpha^2} \Bigr ( \fr{1}{2}\fr{B''}{B}-\fr{1}{4}\fr{A'}{A}\fr{B'}{B}+\fr{1}{r}\fr{B'}{B}-\fr{3}{8} \left (\fr{B'}{B}\right )^2 \Bigr )+\fr{1}{8r} \Bigl ( -\fr{A'}{A}-\fr{A}{r}+\fr{1}{r} \Bigr ) \Bigr ]=0 .\label{eq:ueom_t3}
\eea 
Note that in \eq{eq:ueom_t3} any derivatives with respect to $t$ are canceled out and only the spacial derivative with respect to $r$ remains.
From Eqs.~(\ref{app:A_V}), (\ref{eq:dB}), (\ref{eq:ddB}) and (\ref{eq:dA}) make the left hand side of \eq{eq:ueom_t3} identically vanishes. Thus our solution \eq{eq:B_V}, \eq{app:A_V} and \eq{eq:l_V} satisfies the $t-$ component of the field equation \eq{eq:ueom_t}.

We proceed to the $r-$component of \eq{eq:ueom},
\bea
 \fr{1}{a+1}\bigl [ D_\alpha A^{\alpha r} +2R^{r}_{~\alpha}u^\alpha \Bigr ] +\frac{32\pi G}{c^4}\lambda u^r =0 \label{eq:ueom_r}
\eea
Noting that
\bea
	\fr{1}{a+1}g^{rr}g^{tt}A_{tr}+2g^{rr}R_{rt}u^t
=
	\fr{1}{a+1}\fr{1}{A\sqrt{B}}\left (-\fr{\dot{B}'}{2B}+\fr{\dot{B}B'}{2B^2}+\fr{1}{4}\fr{\dot{A}}{A}\fr{B'}{B}+\fr{2}{r}\fr{\dot{A}}{A} \right )=0\;, \label{eq:ueom_r3}
\eea
we see that \eq{eq:ueom_r3} holds using \eq{eq:dotA} and $\dot{B}=0$. 

Both the $\theta -, \phi-$ components of the field equation,
\bea
 &&\fr{1}{a+1}\bigl [ D_\alpha A^{\alpha \theta} +2R^{\theta}_{~\alpha}u^\alpha \Bigr ] +\frac{32\pi G}{c^4}\lambda u^\theta =0\;, \\
 &&\fr{1}{a+1}\bigl [ D_\alpha A^{\alpha \phi} +2R^{\phi}_{~\alpha}u^\alpha \Bigr ] +\frac{32\pi G}{c^4}\lambda u^\phi =0 \;
\eea
 can be checked by observing that the first order covariant derivative and the Ricci tensor $R_{\theta t}$ and $R_{\phi t}$ are identically zero and that the last terms proportional to $\lambda$ are zero. Thus we conclude that our solution $B(V),A(V),\lambda(V)$ satisfies all the components of the field equation of $u^\mu$.


%
%
%
\end{document}